\documentclass[11pt,a4paper]{article}
\usepackage[utf8]{inputenc}
\usepackage[T1]{fontenc}
\usepackage{amsmath,amssymb,amsthm,amsfonts}
\usepackage{graphicx}
\usepackage{booktabs}
\usepackage{tabularx}
\usepackage{setspace,geometry}
\usepackage[dvipsnames]{xcolor}
\usepackage{cite}
\usepackage{hyperref}
\usepackage{cleveref}

\graphicspath{{./}}

\newcolumntype{L}{>{\raggedright\arraybackslash}X}
\newcommand{\kt}{k_BT}

\title{\textbf{A General Theory for Phenotypic Association in Biological Systems}\\[6pt]}
\author{Giuseppe Battaglia$^{1,2*}$ \\
	\small $^1$Catalan Institution for Research and Advanced Studies (ICREA), Barcelona, Spain. \\
	\small $^2$Institute for Bioengineering of Catalonia (IBEC), \\
	\small The Barcelona Institute of Science and Technology (BIST), Barcelona, Spain.}
\date{}

\begin{document}
\maketitle

\begin{abstract}
\noindent Biological recognition rarely rests on one strong bond. It works by forming many weak ones at once, between crowded, deformable surfaces in water. This review develops that process as a problem in statistical mechanics. Counting the ways two multivalent objects can bind proves to be the classical monomer–dimer problem on a graph, with a rigorous consequence: the apparent switching of multivalent binding is always a smooth crossover, never a phase transition. Three constraints follow. A repulsive surface layer is obligatory rather than a design choice; bonds do not act independently; and since free energies enter rates exponentially, small changes in receptor number shift binding lifetimes by orders of magnitude. One set of equations then covers antibodies, lipoproteins and T cell recognition. In each, what decides the outcome is not the strength of any single bond but how a fixed total is spread over many: affinity is a property of a molecule, selectivity a property of an assembly.
\end{abstract}

\tableofcontents

\part*{\centering Part I \quad Foundations: Two Multivalent Units in Water}
\addcontentsline{toc}{section}{\textbf{Part I. Foundations: Two Multivalent Units in Water}}

\section{Introduction}

Life is an emergent phenomenon. It arises from the collective behaviour of an astronomical number of interacting parts and, this is the remarkable thing, from a startlingly short list of ingredients. Twenty amino acids, four nucleotides, a few dozen lipids and sugars. That is nearly the whole inventory. Yet the combinatorial assembly of these monomers generates a universe of macromolecules whose structural and chemical diversity defeats any attempt at exhaustive enumeration. Every protein fold, every nucleic-acid sequence, every multicomponent complex is, in practice, unique, and the information encoded in such structures grows factorially as the system grows in size and connectivity. This combinatorial explosion is what endows living systems with their extraordinary capacity for information storage, adaptive response and functional specialisation.

But the very richness that makes life possible also imposes a punishing set of physical constraints. The cell is not a dilute solution of isolated molecules politely awaiting their partners. It is a crowded metropolis, where macromolecular concentrations routinely exceed $300$~mg\,mL$^{-1}$ \cite{Zimmerman1991,Ellis2001,Zhou2008}, where every object is continuously buffeted by thermal noise, and where each entity is simultaneously engaged in a dense network of specific and non-specific encounters. A protein in the cytoplasm is never alone, never still, and never more than a nanometre or two from something it could stick to. The wonder is not that recognition sometimes fails. The wonder is that it works at all.

To understand how order is maintained and function executed in such a place, one must look at biological matter through the lens of statistical physics. The forces governing this molecular society divide into two opposing classes, and the distinction is worth stating carefully because everything downstream depends on it.

The first class is \emph{attraction}. These interactions are specific, directional and typically short-ranged, arising from the precise complementarity of shape, charge and hydrophobicity across a molecular interface. They are what allows an enzyme to pick its substrate out of a sea of metabolites, a transcription factor to find its cognate motif among billions of base pairs \cite{Berg1981}, a signalling protein to dock on its receptor and no other. When specificity and directionality are preserved, attraction drives the formation of well-defined, functional assemblies. But attraction carries an inherent risk. If binding interfaces become promiscuous, through mutation, denaturation, or simply an increase in concentration, the same forces drive uncontrolled oligomerisation and, eventually, catastrophic phase transitions. Biology exploits a controlled version of this: liquid--liquid phase separation into membraneless organelles is attraction operating deliberately at the edge of condensation \cite{Brangwynne2009,Nott2015,Banani2017}. Amyloid aggregation in neurodegeneration is the same physics with the brakes off, attraction that has stopped discriminating \cite{Chiti2006,Chiti2017,Knowles2014}.

The second class is \emph{repulsion}. These interactions are non-specific, roughly isotropic, and prevent objects from approaching too closely: excluded volume, electrostatic double layers, and the steric haze thrown up by disordered protein segments and surface glycans \cite{Weinbaum2007,Mockl2020}. Repulsive forces are the guardians of biological identity. In the absence of a correct partner they hold each macromolecule dispersed and soluble, poised to engage when the right partner appears. Strip them away and the crowded cytoplasm would collapse into a single functionless aggregate within seconds. It is the balance between the two classes, attraction conferring the ability to form specific complexes and repulsion preserving the individuality of each component, that underpins the hierarchical organisation of living matter, from protein complexes through cellular compartments to the architecture of tissues.

Here we should be honest about something that is easy to gloss over. Neither class of force acts in a vacuum. Both act in liquid water, at body temperature, in salt, surrounded by crowders. And water is not a passive stage on which the interesting physics is performed. Water is a participant. The hydrophobic effect is not a force between oily surfaces at all; it is the water between them, paying an entropic penalty for staying ordered, and being released \cite{Chandler2005,Southall2002}. Hydration forces oscillate with a period of a quarter of a nanometre because that is how wide a water molecule is \cite{Israelachvili1996,Leikin1993}. Screening shrinks the reach of the electrostatic interaction from micrometres to under a nanometre. Every interaction we care about is really a \emph{solvent-averaged} interaction, a free energy rather than an energy, and, as we shall see, that single fact is what makes the whole problem a many-body problem rather than a sum of pairs.

There is a further complication, which becomes the subject of Part~II. Living systems do not sit at equilibrium. Prigogine and co-workers showed that ordered structures emerge spontaneously in systems held far from equilibrium by a continuous flux of energy and matter \cite{Prigogine1977}, and England extended the argument to show that groups of molecules can organise into configurations that dissipate energy more effectively \cite{England2015}. The attractive forces that drive biological assembly are not static; they are sculpted by the metabolic free-energy flux. The resulting non-equilibrium steady state is exquisitely sensitive to perturbation and can display the hallmarks of self-organised criticality \cite{Bak1987} and deterministic chaos \cite{Lorenz1963,Strogatz2015}. That sensitivity enables flexible adaptation, but it carries a long-term cost: the slow, irreversible drift of control parameters, such as protein oxidation, matrix cross-linking and epigenetic noise, that we experience as ageing \cite{LopezOtin2013}.

The theoretical challenge, then, is to build a description that captures the interplay of specific attraction and non-specific repulsion across the enormous range of scales biology spans, from the nanometre protein interface to the micrometre cell, while respecting the liquid, crowded, thermally agitated and ultimately non-equilibrium nature of the medium. 

\section{The Medium: Water, and Why Interactions Are Free Energies}\label{sec:medium}

\subsection{An honest inventory of the medium}

Before writing a single interaction down, it pays to look at the numbers describing the place where it happens. \Cref{tab:medium} collects them, all computed from fundamental constants rather than quoted.

Three entries deserve to be read twice. The Debye length at physiological ionic strength is $0.785$~nm. Electrostatics, the textbook long-range interaction, the one that reaches across a room in vacuum, is inside a cell a contact interaction. It reaches about two water molecules. Second, the thermal energy $\kt$ is $2.5$~kJ\,mol$^{-1}$, which sets the currency for everything: an interaction worth $1\,\kt$ is a suggestion, one worth $20\,\kt$ is a life sentence. Third, and this is the entry that licenses the entire theoretical apparatus that follows, water rearranges in a few picoseconds while a $5$-nm object needs about $10$ nanoseconds to shuffle a single nanometre. That is a factor of ten thousand.

\begin{table}[t!]
\centering
\small
\caption{The physiological aqueous medium at $T=298$~K, computed from fundamental constants. $I$ denotes ionic strength.}
\label{tab:medium}
\begin{tabularx}{\textwidth}{@{}l l l L@{}}
\toprule
\textbf{Quantity} & \textbf{Symbol} & \textbf{Value} & \textbf{Why it matters} \\
\midrule
Thermal energy & $\kt$ & $4.12$~pN\,nm & the currency of all free energies \\
 & & $2.48$~kJ\,mol$^{-1}$ & $=0.593$~kcal\,mol$^{-1}=25.7$~meV \\
Relative permittivity & $\varepsilon_r$ & $78.4$ & Coulomb forces cut $\sim$80-fold \\
Bjerrum length & $\lambda_B$ & $0.715$~nm & where charge interaction $=\kt$ \\
Debye length, $I=10$~mM & $\kappa^{-1}$ & $3.04$~nm & in vitro, electrostatics reaches \\
Debye length, $I=150$~mM & $\kappa^{-1}$ & $0.785$~nm & \textbf{in vivo, electrostatics is contact-range} \\
Debye length, $I=1$~M & $\kappa^{-1}$ & $0.304$~nm & essentially switched off \\
Hydration decay & $\lambda_w$ & $\approx0.3$~nm & oscillates with period $0.25$~nm \\
Hydrophobic term & $\gamma_{\rm eff}$ & $\approx2.5\,\kt$\,nm$^{-2}$ & area-driven, entropic in origin \\
Viscosity & $\eta$ & $0.89$~mPa\,s & motion is overdamped, no inertia \\
Diffusivity, $R=5$~nm & $D$ & $49\ \mu$m$^2$\,s$^{-1}$ & $\sim$10~ns to move 1~nm \\
Solvent relaxation & $\tau_w$ & $1$--$5$~ps & $10^3$--$10^4\times$ faster than the unit \\
\bottomrule
\end{tabularx}
\end{table}

\subsection{Averaging over the water}

Suppose we wanted to be scrupulous. We would write down the coordinates of the two objects whose meeting interests us, and then the coordinates of everything else: every water molecule, every ion, every metabolite, and every wobble of the objects' own internal structure. The exact interaction may well be a tidy sum of pairwise atomic terms. But nobody can solve that, and more importantly nobody needs to. What governs the association, and what an experiment measures, is the interaction \emph{averaged over all those other coordinates}.

Let $\mathbf{R}$ collect the coordinates we care about, the positions and orientations of the interacting objects, and let $\mathbf{r}$ collect everything else. Integrating out $\mathbf{r}$ at fixed $\mathbf{R}$ defines the potential of mean force \cite{Kirkwood1935,Roux1995},
\begin{equation}\label{eq:pmf}
W(\mathbf{R})=-\kt\,\ln\!\int\! \mathrm{d}\mathbf{r}\;e^{-\beta U(\mathbf{R},\mathbf{r})}+\text{const},
\qquad \beta\equiv1/\kt,
\end{equation}
whose gradient $-\nabla_\mathbf{R}W$ is the mean force acting along the retained coordinates. Three consequences of \cref{eq:pmf} shape everything that follows.

The first is a matter of vocabulary, but it is not pedantry. $W$ is a \emph{free energy}: it contains entropy, depends on temperature, and may be entirely entropic in origin, as it is for the hydrophobic effect and for depletion. We shall therefore write bond \emph{free energies} $\varepsilon$ throughout and never bond energies. When a biologist reports a binding energy of $-10\,\kt$ and a physicist writes that number into a potential, they are talking about different objects unless both remember that the number already contains the entropy of released water, reorganised side chains and displaced counterions. Its enthalpic and entropic parts are not fixed, and they respond differently to temperature and to mutation.

The second consequence is what makes the theory tractable. Equation~\eqref{eq:pmf} is only a useful dynamical object if the integrated-out coordinates equilibrate quickly compared with the motion of $\mathbf{R}$, and \cref{tab:medium} shows that they do, by four orders of magnitude. The solvent is therefore always at equilibrium with respect to the instantaneous configuration of the units, and $W(\mathbf{R})$ acts as a genuine potential governing their motion. The same estimate shows that the Reynolds number is minuscule and inertia irrelevant: nothing at this scale coasts. Motion is overdamped, which is exactly why Part~II describes the dynamics with a Smoluchowski equation rather than with Newton's.

The third consequence is the one to circle, because it is the seed of \cref{sec:manybody}. \emph{The logarithm in \cref{eq:pmf} does not distribute over a sum.} Even if $U(\mathbf{R},\mathbf{r})$ is strictly pairwise additive, $W(\mathbf{R})$ is in general not. Many-body forces are not an exotic correction that fastidious theorists add at the end; they are generated automatically, by the mere act of working in a liquid instead of a vacuum. A theory of association in water that assumes forces add is not simplified, it is wrong by an amount we shall shortly compute.

\section{The Alphabet of Forces}\label{sec:alphabet}

\subsection{Four questions to ask of any interaction}

\begin{figure}[t!]
\centering
\includegraphics[width=\textwidth]{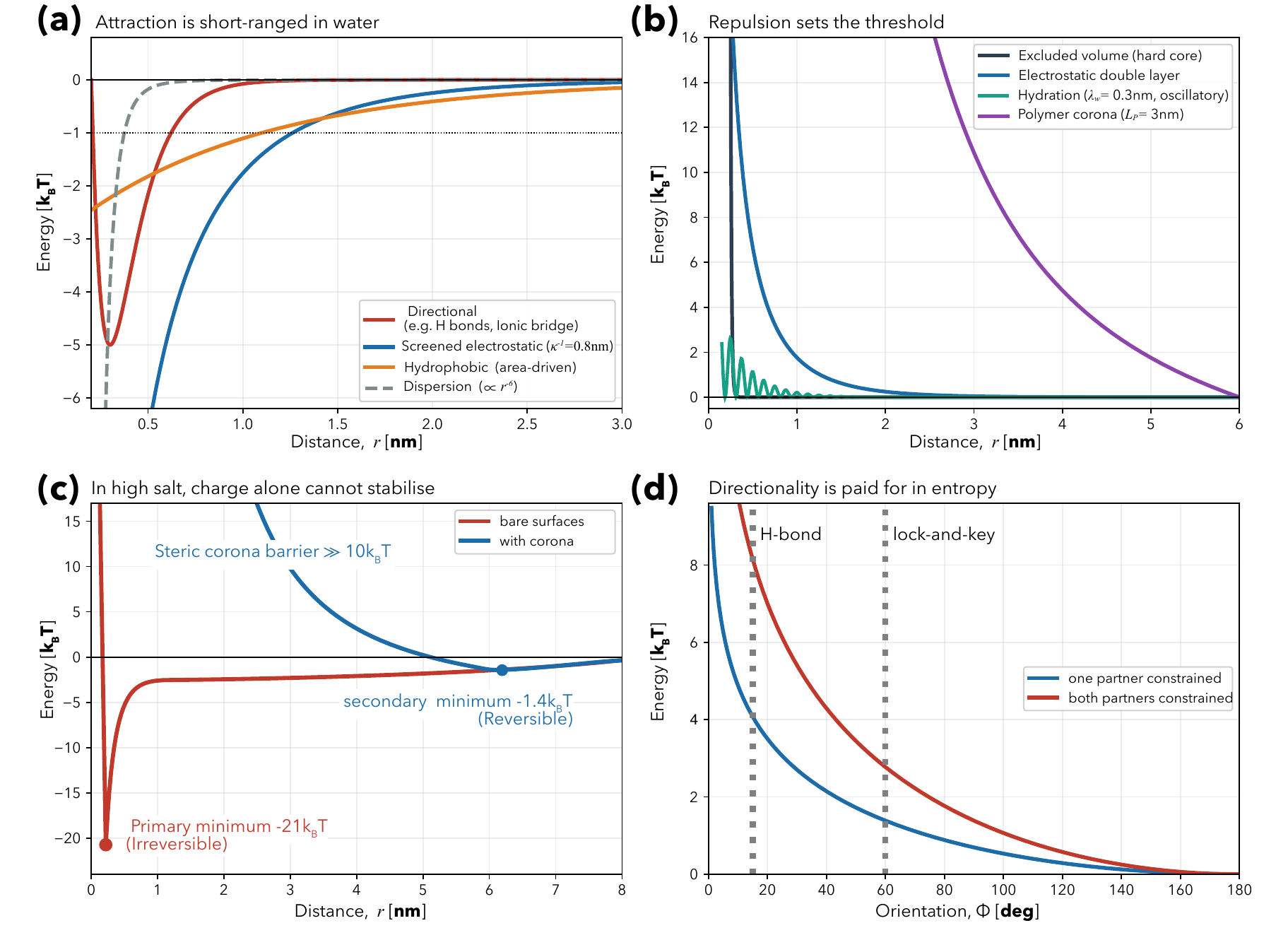}
\caption{The elementary contributions to the pair potential of mean force in water, all in units of $\kt$. \textbf{(a)}~The specific, directional terms are contact-ranged; the non-specific ones are weak or area-driven. \textbf{(b)}~Repulsive terms: a hard core, a double layer screened to $0.79$~nm, hydration oscillating with the width of a water molecule, and a polymer corona that dominates out to twice its height. \textbf{(c)}~The net PMF between two units at physiological salt. Bare surfaces have \emph{no} barrier anywhere and fall into a well $21\,\kt$ deep; a steric corona raises a barrier far above $10\,\kt$ and leaves only a shallow, reversible secondary minimum. \textbf{(d)}~The entropic bill for directionality, \cref{eq:orient}.}
\label{fig:potentials}
\end{figure}

Biology builds recognition out of a modest set of physical interactions, and one can characterise any of them by asking four questions. How strong is it? How far does it reach? How picky is it about \emph{which} partner? And how picky is it about the partner's \emph{orientation}? The first two are familiar; the second two are usually left qualitative, and we will fix that.
For moieties $a$ and $b$ at separation $\mathbf{r}$ with orientations $\Phi_a,\Phi_b$, write the elementary pair potential of mean force in the separable form
\begin{equation}\label{eq:separable}
u_{ab}(\mathbf{r},\Phi_a,\Phi_b)=
\underbrace{\varepsilon_{ab}}_{\text{strength}}\;\times\;
\underbrace{f(r/\lambda)}_{\text{range}}\;\times\;
\underbrace{g(\Phi_a,\Phi_b)}_{\text{directionality}},
\end{equation}
with $\varepsilon_{ab}$ the depth at optimal geometry, $\lambda$ the characteristic decay length, and $0\le g\le1$ an angular kernel. Representative radial forms are dispersion, $f\propto(\sigma/r)^{6}$; the screened Coulomb interaction between charges $z_a,z_b$,
\begin{equation}\label{eq:dh}
u=\frac{\lambda_B z_az_b}{r}\,e^{-\kappa r}\,\kt ;
\end{equation}
the area-driven hydrophobic term $u=-\gamma_{\rm eff}A_{\rm c}e^{-h/\lambda_h}$; and the oscillatory hydration term $u=A_we^{-h/\lambda_w}\cos^2(2\pi h/d_w)$. The factorisation in \cref{eq:separable} is an approximation, since real interactions couple distance and angle, but it is the right organising skeleton because it makes each descriptor separately measurable and separately designable.
\Cref{tab:interactions} answers the four questions for the whole alphabet, drawing on the standard treatments of surface forces \cite{Israelachvili2011,Parsegian2006}, and \cref{fig:potentials}(a,b) plots the answers. Two features of that table organise everything that follows.

The first is that the strong interactions are useless for recognition. A covalent bond is worth $150$ to $400\,\kt$; on any biological timescale it is permanent. Biology uses such bonds to \emph{build} objects, not to recognise them. Reversible recognition lives in a narrow window, roughly $1$ to $15\,\kt$, strong enough to be selective and weak enough to let go. Everything interesting happens in that band.

The second is that every attractive interaction in that band is short-ranged \emph{in water}. Dispersion dies as the sixth power of distance. Hydrogen bonding is effectively a contact phenomenon. And electrostatics, which would reach far in vacuum, is screened to $0.79$~nm. There is no long-range specific attraction available to a cell. Specific recognition\textit{ in vivo} is something that happens on contact or not at all, which immediately raises the question of how anything ever finds anything, and foreshadows why the non-specific, longer-ranged repulsive terms end up controlling the whole process.

\begin{table}[t!]
\centering
\small
\caption{The alphabet of interactions available to biological matter in water. Strengths are per interaction at optimal geometry; $\phi_0$ is the angular tolerance. Specificity is the cognate versus non-cognate contrast $\Delta\varepsilon$ of \cref{eq:specificity}.}
\label{tab:interactions}
\setlength{\tabcolsep}{4pt}
\begin{tabularx}{\textwidth}{@{}L c l l l l@{}}
\toprule
\textbf{Interaction} & \textbf{Sign} & \textbf{Strength} $[\kt]$ & \textbf{Range} & \textbf{Direction.} & \textbf{Specificity}\\
\midrule
Covalent bond            & $-$ & $150$--$400$ & $0.15$~nm & very high & very high\\
Metal coordination       & $-$ & $20$--$80$   & $0.2$~nm  & high      & high\\
Hydrogen bond            & $-$ & $2$--$10$    & $0.3$~nm  & high, $\sim15^\circ$ & high\\
Salt bridge              & $-$ & $1$--$5$     & $\kappa^{-1}$ & moderate & moderate\\
$\pi$--$\pi$, cation--$\pi$ & $-$ & $1$--$4$  & $0.4$~nm  & moderate  & moderate\\
Dispersion (van der Waals) & $-$ & $0.2$--$2$ & $r^{-6}$  & none      & none\\
Hydrophobic              & $-$ & $2.5$ per nm$^2$ & $\sim1$~nm & none (area) & low\\
\midrule
Excluded volume          & $+$ & $\rightarrow\infty$ & $<0.3$~nm & none & none\\
Electrostatic double layer & $+$ & $1$--$20$ & $\kappa^{-1}$ & none & none\\
Hydration                & $+$ & $1$--$10$  & $0.3$~nm, oscill. & none & none\\
Polymer corona, brush\,\cite{Alexander1977,deGennes1987,Milner1988} & $+$ & $10$--$10^2$ & $2L$ & none & none\\
Membrane undulation\,\cite{Helfrich1973} & $+$ & $\sim1$ & $\sim$nm & none & none\\
Configurational entropy loss & $+$ & $1$--$10$ & n/a & n/a & n/a\\
\bottomrule
\end{tabularx}
\end{table}

\subsection{Directionality has a price, and it is steep}

A directional interaction demands that its partner arrive not merely close but correctly oriented. That demand costs entropy, because it throws away most of the orientations the partner was previously free to explore, and the bookkeeping is elementary. If a moiety must present itself within a cone of half-angle $\phi_0$ about a preferred axis, the accessible fraction of solid angle is
\begin{equation}\label{eq:solidangle}
\frac{\int_0^{\phi_0}\sin\theta\,\mathrm{d}\phi}{\int_0^{\pi}\sin\phi\,\mathrm{d}\phi}
=\frac{1-\cos\phi_0}{2},
\end{equation}
and since the free energy of a restriction is minus $\kt$ times the logarithm of the surviving fraction of configuration space, forming the interaction costs
\begin{equation}\label{eq:orient}
\Delta G_{\rm orient}=-\kt\ln\!\left(\frac{1-\cos\phi_0}{2}\right)
\qquad\text{per constrained partner.}
\end{equation}
If rotation about the axis is also restricted to a range $\Delta\phi$, a further $-\kt\ln(\frac{\Delta\phi}{2\pi})$ must be added, and for two independently constrained partners the penalties simply add because the restrictions are on independent degrees of freedom.

\Cref{fig:potentials}(d) plots \cref{eq:orient}, and the numbers are sobering. A hydrogen bond requires $\phi_0\approx15^\circ$, which costs $4.1\,\kt$ per partner and $8.1\,\kt$ if both partners are pinned. A lock-and-key interface with $\phi_0\approx60^\circ$ costs $1.4\,\kt$ per partner. An isotropic interaction costs nothing. Now compare with what the bond returns: a hydrogen bond delivers between $2$ and $10\,\kt$. The bond can therefore be, on balance, barely favourable, or even unfavourable, despite being individually the most reliable specific interaction in the biological toolkit.

This is not a defect of the accounting; it is a real and rather beautiful constraint, and it explains three things at once. It explains why single specific contacts are so rarely used alone. It explains why multivalency is not a clever optimisation but very nearly a requirement: when many contacts form between two objects already brought into register, the orientational bill is paid \emph{once}, by the first contact, and amortised over all the rest. And it explains why rigid, pre-organised scaffolds outperform floppy ones at equal valency, since a rigid scaffold has paid its orientational entropy during synthesis rather than during binding. We will meet all three consequences again, the last of them quantitatively in \cref{sec:topology}.

\subsection{Specificity is a property of a matrix, not of a bond}

It is tempting to say that a strong interaction is a specific one. It is not. Specificity is about \emph{contrast}, that is, how much better the cognate partner does than the alternatives. A very strong interaction that binds everything equally well is perfectly non-specific, and a weak one that binds one partner slightly better than all others can be exquisitely selective given enough repetitions.

Index moiety types $a=1\ldots d$ on one object and $b=1\ldots k$ on the other, and collect the strengths into a matrix $\boldsymbol{\varepsilon}\in\mathbb{R}^{d\times k}$. For a cognate partner $b^\star$ the relevant quantity is the free-energy contrast against the non-cognate background,
\begin{equation}\label{eq:specificity}
\Delta\varepsilon_a=\bigl\langle\varepsilon_{ab}\bigr\rangle_{b\neq b^\star}-\varepsilon_{ab^\star},
\end{equation}
and because binding constants are exponential in free energy, the discrimination ratio between cognate and typical non-cognate binding is
\begin{equation}\label{eq:discrim}
\frac{K_{ab^\star}}{\langle K_{ab}\rangle}=e^{\beta\,\Delta\varepsilon_a}.
\end{equation}
The scale is worth committing to memory: $\Delta\varepsilon=2.3\,\kt$ buys one decade of discrimination, $4.6\,\kt$ buys two, $9.2\,\kt$ buys four. Non-specific interactions, meaning dispersion, hydrophobic contact, excluded volume and the double layer, have $\Delta\varepsilon\approx0$ \emph{by construction}, since they act on all partners alike.

Equation~\eqref{eq:discrim} is the formal reason the attraction and repulsion dichotomy of the Introduction is more than a taxonomy. Attraction can carry information, because its matrix has contrast. Repulsion cannot, because its matrix is flat. What repulsion can do instead, and this turns out to be at least as important, is set a \emph{threshold}: a uniform toll that every partner must pay, which the cognate partner can afford only by summing many small attractive contributions. Information from attraction, gating from repulsion. Keep the division in mind, because everything that follows turns on it.

\section{Why Repulsion Is Not Optional}\label{sec:necessity}

We can now establish something concrete and slightly alarming. It is standard to think of charge as what keeps biological objects apart: proteins have isoelectric points, membranes are anionic, DNA is a polyelectrolyte. Let us simply add up the terms and see whether charge is up to the job at physiological salt.
\Cref{fig:potentials}(c) superposes dispersion, screened double-layer repulsion, hydration and a hard core for two objects of radius $20$~nm at $I=150$~mM. The result is unambiguous, and it follows from a competition of functional forms rather than from any particular parameter choice. Dispersion between two spheres falls off algebraically, roughly as $R/h$ \cite{Hamaker1937,Parsegian2006}. The double layer falls off exponentially, with decay length $0.785$~nm. An exponential always loses to a power law eventually, and here ``eventually'' means ``almost immediately''. The consequence is that the net potential of mean force has \emph{no repulsive barrier above zero at any separation}. It descends monotonically into a primary minimum of $-20.7\,\kt$ at contact, a failure of the classical stabilisation picture \cite{Derjaguin1941,Verwey1948}.

How deep is $21\,\kt$? Escape requires a fluctuation of that size, which happens with probability $e^{-21}\approx10^{-9}$. Two bare surfaces of this size that touch in physiological buffer are joined for good. Charge repulsion, at the salt concentration life actually runs at, does not stabilise anything.

Now add a grafted polymer corona of height $3$~nm made of hydrophilic neutral motifs, capabale of attracting water but effectively repelling anything else. The picture changes qualitatively rather than quantitatively. The corona contributes an osmotic and elastic repulsion out to twice its height, raising a barrier far in excess of $10\,\kt$, and the deepest remaining feature is a shallow secondary minimum of $-1.4\,\kt$ sitting at the corona edge. That is comparable to thermal energy: objects can visit, and objects can leave.

This deserves to be stated as a structural conclusion rather than a modelling observation. \emph{An object that is to recognise something must first be prevented from sticking to everything, and in water at physiological salt this cannot be achieved with charge.} It requires a steric, non-specific, entropically generated repulsive layer. Biology's coronas are therefore not decoration: the glycocalyx, the disordered tails of membrane proteins, surface glycans, the mucins. They are the thermodynamic precondition that makes selective recognition possible at all. Synthetic constructs inherit exactly the same requirement, which is why poly(ethylene glycol) is not a finishing touch \cite{Jokerst2011,Suk2016}. It is also why the protein corona that assembles on a nanoparticle in plasma so often abolishes its targeting: the biological identity of the construct is set by the layer it acquires, not by the layer it was given \cite{Monopoli2012,Walkey2012,Salvati2013}. And notice the pleasing corollary: because the corona is what holds the two objects apart at a well-defined distance, it also decides how many attractive contacts can form. The gatekeeper sets the terms of the negotiation. We cash that in shortly.

\section{The Unit: Defining the Agent}\label{sec:unit}

We need a name for the kind of object whose interactions we are describing, and it should be general enough to cover a virus, an antibody, a lipoprotein, a nanoparticle, a polymer and a patch of cell membrane, because we intend to treat all of them with the same equations. We define a \emph{multivalent unit} as the triple
\begin{equation}\label{eq:unit}
\mathcal{U}=\bigl(\mathcal{S},\;\{\mathcal{L}_\zeta\}_{\zeta=1}^{T},\;\mathcal{C}\bigr),
\end{equation}
a \emph{scaffold} $\mathcal{S}$ carrying $T$ populations of binding \emph{moieties} $\mathcal{L}_\zeta$, wrapped in a repulsive \emph{corona} $\mathcal{C}$, and we characterise it by the descriptor set
\begin{equation}\label{eq:descriptors}
\boldsymbol{\Lambda}=\bigl\{\;\ell_\zeta,\;\varepsilon_\zeta,\;\phi_{0,\zeta},\;\mathbf{g},\;\xi,\;L,\;\sigma_c,\;R\;\bigr\},
\end{equation}
where $\ell_\zeta$ is the valency of type $\zeta$; $\varepsilon_\zeta$ and $\phi_{0,\zeta}$ its strength and angular tolerance from \cref{eq:separable,eq:orient}; $\mathbf{g}$ the spatial arrangement of the moieties, meaning their mean spacing and whether that spacing is ordered or random; $\xi$ the scaffold compliance, running from rigid ($\xi\to0$: DNA origami \cite{Rothemund2006}, a viral capsid) to freely flexible ($\xi\to\infty$: a random coil); $L$ and $\sigma_c$ the corona height and grafting density; and $R$ the overall size.

What makes an object a unit is not its chemistry but a topological fact: it presents several binding moieties on a \emph{shared body}, so that binding events are coupled through the scaffold. Bind one moiety and you have moved all the others. This is the whole source of both the power and the difficulty of multivalency, and we will see it return in \cref{sec:manybody} as an irreducible many-body term that no amount of care with pair potentials can capture.

Of the descriptors in \cref{eq:descriptors}, three will do most of the work. \emph{Valency} $\ell_\zeta$ sets the combinatorial scale. \emph{Compliance} $\xi$ arbitrates the trade-off we met in \cref{eq:orient}: a rigid unit pays its orientational entropy once, at synthesis, but must then match the target's geometry, whereas a flexible unit adapts to any geometry but pays a conformational penalty for every bond it forms. And the \emph{corona} $(L,\sigma_c)$ sets the threshold of \cref{sec:necessity} and, as promised, also the ceiling on engagement.

Put two units in the medium of \cref{sec:medium} and ask what we need to track. Remarkably little, as it turns out. The units are described by the separation $h$ between their surfaces, or the contact area $A_c$ for deformable units, together with the \emph{engagement vector}
\begin{equation}\label{eq:engagement}
\mathbf{k}=(k_1,\ldots,k_T),
\end{equation}
whose entries count the bonds formed by each moiety type. Slow internal coordinates, meaning corona compression, scaffold strain and partner redistribution within the contact, may be appended where needed.

The asymmetry between the two coordinates is the technical heart of the problem. Separation $h$ is one continuous number. Engagement $\mathbf{k}$ is discrete, high-dimensional and, crucially, each value of it can be realised in an enormous number of ways. Counting those ways correctly is the subject of the next section, and it is not bookkeeping for its own sake: the count enters the free energy logarithmically, is often worth many $\kt$, and is the reason multivalent recognition can switch on far more sharply than any single interaction does.

\section{The Topology Kernel: Counting the Ways to Bind}\label{sec:topology}

\subsection{Engagement as a matching problem}

The number of distinct microstates that realise a given engagement $\mathbf{k}$ is the \emph{topology kernel}
\begin{equation}\label{eq:topology}
\Omega(\mathbf{k})=\#\{\text{microstates realising }\mathbf{k}\},
\qquad S_{\rm comb}=k_B\ln\Omega(\mathbf{k}),
\end{equation}
and the combinatorial entropy $S_{\rm comb}$ it defines is a genuine free-energy contribution, not a normalisation. It is the term that Kitov and Bundle isolated as the statistical origin of the multivalency effect \cite{Kitov2003}, and the term whose growth with receptor density produces superselectivity \cite{Martinez-Veracoechea2011}.
Writing $\Omega$ down for a particular architecture is usually treated as a modelling choice. It is not: it is a well-posed counting problem with a name. Represent the encounter as a bipartite \emph{reachability graph} $G=(\mathcal{L}\cup\mathcal{R},E)$, whose left vertices are the $\ell$ ligands of the unit, whose right vertices are the $R$ receptors lying within the contact, and which contains the edge $(i,j)\in E$ if and only if ligand $i$ can physically reach receptor $j$ given the scaffold geometry, the spacer length and the corona-imposed separation. A microstate with $k$ bonds is then a set of $k$ edges of $G$ that share no vertex, because a ligand cannot bind twice and a receptor cannot be bound twice. Such a set is precisely a \emph{$k$-matching}, and therefore
\begin{equation}\label{eq:matching}
\boxed{\;\Omega(k)=m_k(G),\;}
\end{equation}
the number of $k$-matchings of the reachability graph. Every architectural question, ligand spacing, spacer reach, receptor order, scaffold rigidity and steric blocking, enters the thermodynamics through the edge set $E$ and nowhere else.

This identification is useful for three reasons. It supplies closed forms in the cases that matter, which we give in \cref{sec:arrangements}. It connects the problem to a solved area of statistical mechanics, the monomer--dimer problem, from which we can import a rigorous theorem about the sharpness of multivalent thresholds (\cref{sec:heilmann}). And it tells us immediately when the counting is hard: computing $m_k$ for a general graph is equivalent to evaluating a permanent, which is \#P-complete \cite{Valiant1979}, so exact enumeration is available only for structured or small graphs and controlled approximation is needed otherwise \cite{Kasteleyn1967,MezardMontanari2009}.

\subsection{Five arrangements and their kernels}\label{sec:arrangements}

\begin{figure}[t!]
\centering
\includegraphics[width=\textwidth]{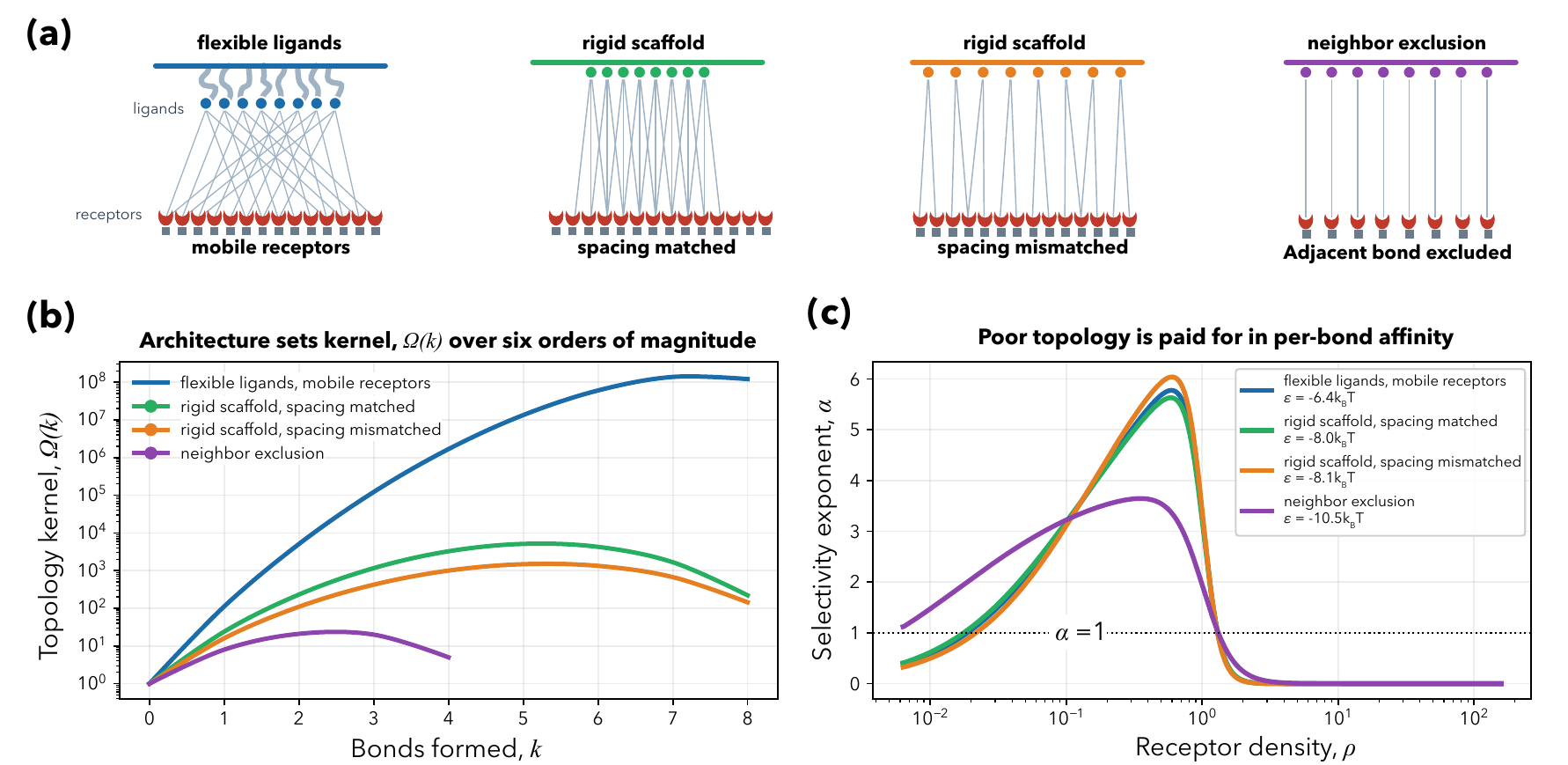}
\caption{The topology kernel is a $k$-matching count of the bipartite reachability graph, \cref{eq:matching}. \textbf{(a)}~Reachability graphs for four architectures with $\ell=8$ ligands (circles, on the scaffold) and $R=14$ receptors (squares); an edge means the ligand can reach the receptor. \textbf{(b)}~The resulting kernels, computed exactly, spanning six orders of magnitude at fixed valency. \textbf{(c)}~Selectivity exponent $\alpha$ when the per-bond free energy is chosen so that all four architectures place their threshold at the same receptor density. The achievable sharpness is similar in the first three cases; what a poor topology costs is per-bond affinity, quoted in the legend. Neighbour exclusion is different in kind, because it removes states outright and caps $\alpha$.}
\label{fig:topology}
\end{figure}

\Cref{fig:topology} shows four reachability graphs and the kernels they generate; \cref{tab:omega} collects the closed forms, including a fifth case treated below. All counts quoted are exact, obtained by dynamic programming over receptor occupancy and validated against the closed forms where these exist.

\paragraph{Flexible ligands, mobile receptors.} If every ligand can reach every receptor, $G$ is the complete bipartite graph $K_{\ell,R}$ and the count factorises into three independent choices: which $k$ of the $\ell$ ligands bind, which $k$ of the $R$ receptors they bind, and which of the $k!$ pairings between the two chosen sets is realised. Hence
\begin{equation}\label{eq:omegaflex}
\Omega^{\rm flex}(k)=\binom{\ell}{k}\binom{R}{k}\,k!\,.
\end{equation}
This is the standard result, appropriate to a flexible polymer bearing pendant ligands against a fluid membrane in which receptors diffuse freely, and it is the kernel used in nearly all existing superselectivity theory. It is also the most generous kernel possible at fixed $\ell$ and $R$, since adding edges to a graph can only increase its matching numbers.

\paragraph{Rigid scaffold, commensurate receptors.} A rigid unit presents its ligands at fixed positions $\mathbf{x}_i$, and a ligand tethered by a spacer of reach $\lambda$ can engage only receptors within that distance,
\begin{equation}\label{eq:reach}
(i,j)\in E \iff |\mathbf{x}_i-\mathbf{y}_j|\le\lambda .
\end{equation}
When the ligand spacing matches the receptor spacing, each ligand reaches a small, overlapping set of receptors and $G$ is a sparse band-diagonal graph. The kernel remains large, because the matching can slide along the band, but it is far smaller than \cref{eq:omegaflex}: for the example of \cref{fig:topology} the maximum is $5.1\times10^3$ against $1.4\times10^8$.

\paragraph{Rigid scaffold, incommensurate receptors.} If the ligand spacing is not a near-multiple of the receptor spacing, some ligands fall between receptors and contribute no edges at all. The graph loses vertices of nonzero degree and the kernel falls further, by roughly a factor of three at every $k$ in our example. This is the physical content of the spacing-matching requirement demonstrated with DNA origami by Bastings and co-workers \cite{Bila2022}: geometry does not merely modulate binding, it edits the edge set.

\paragraph{Steric blocking between neighbouring bonds.} Suppose each ligand faces one receptor, but a formed bond sterically occludes the neighbouring sites, as happens for bulky ligands on a close-packed lattice or for large receptors whose footprints overlap. The admissible configurations are then $k$-subsets of $\ell$ sites containing no two adjacent members, a one-dimensional hard-rod problem whose count is
\begin{equation}\label{eq:omegarods}
\Omega^{\rm block}(k)=\binom{\ell-k+1}{k},
\end{equation}
which vanishes for $k>\lceil \ell/2\rceil$. Blocking therefore does something the other architectures do not: it imposes a hard ceiling on engagement that is geometric rather than energetic, independent of how favourable the bonds are.

\paragraph{Tethered ligands, dilute receptors.} A useful limit for design is a flexible tether of reach $\lambda$ against receptors at surface density $\rho$, so that each ligand independently surveys an area $\pi\lambda^2$ containing $n_{\rm eff}=\rho\pi\lambda^{2}$ receptors on average. When $n_{\rm eff}\gg k$ the receptors are effectively never in competition and
\begin{equation}\label{eq:omegatether}
\Omega^{\rm teth}(k)\simeq\binom{\ell}{k}\,n_{\rm eff}^{\,k}
=\binom{\ell}{k}\bigl(\rho\pi\lambda^{2}\bigr)^{k},
\end{equation}
which is the form that generates the compact partition function of \cref{sec:pairwise} and makes the spacer length an explicit design variable: reach enters as $\lambda^{2k}$, so doubling the tether is worth $2k\,\kt\ln 2$ of combinatorial free energy.

\begin{table}[t!]
\centering
\small
\caption{Topology kernels for five arrangements, as $k$-matching counts of the reachability graph. The final column gives the largest $\Omega$ for the worked example of \cref{fig:topology} with $\ell=8$ ligands and $R=14$ receptors.}
\label{tab:omega}
\setlength{\tabcolsep}{4pt}
\begin{tabularx}{\textwidth}{@{}>{\raggedright\arraybackslash}p{3.5cm} L l r@{}}
\toprule
\textbf{Arrangement} & \textbf{Reachability graph} & $\boldsymbol{\Omega(k)}$ & $\boldsymbol{\max_k\Omega}$\\
\midrule
Flexible ligands, mobile receptors & complete bipartite $K_{\ell,R}$ & $\binom{\ell}{k}\binom{R}{k}k!$ & $1.4\times10^{8}$\\
Rigid scaffold, commensurate & sparse band-diagonal & $m_k(G)$, closed form per geometry & $5.1\times10^{3}$\\
Rigid scaffold, incommensurate & sparser, isolated vertices & $m_k(G)$ & $1.5\times10^{3}$\\
Neighbour exclusion & path with adjacency forbidden & $\binom{\ell-k+1}{k}$ & $21$\\
Tethered, dilute receptors & independent reach discs & $\binom{\ell}{k}(\rho\pi\lambda^{2})^{k}$ & set by $\rho\lambda^2$\\
\bottomrule
\end{tabularx}
\end{table}

\subsection{What the matching picture buys: a rigorous statement about sharpness}\label{sec:heilmann}

The identification in \cref{eq:matching} has a consequence that is worth more than the closed forms. Consider the single-type contact partition function that we will construct properly in \cref{sec:pairwise},
\begin{equation}\label{eq:matchpoly}
\Xi(x)=\sum_{k\ge0}\Omega(k)\,x^{k},
\qquad x\equiv e^{-\beta\varepsilon}\;\;(>0),
\end{equation}
which by \cref{eq:matching} is exactly the \emph{matching generating polynomial} of the reachability graph, and therefore exactly the partition function of a monomer--dimer system on that graph with dimer activity $x$. Heilmann and Lieb proved that such a polynomial has only real, non-positive roots, for every graph and every set of non-negative activities, and concluded that a monomer--dimer system cannot undergo a phase transition as a function of monomer density \cite{HeilmannLieb1972}.

Transposed into our language this is a statement about how sharp a multivalent threshold can be, and it cuts in two directions. On the one hand $\Xi(x)$ has no complex zeros approaching the positive real axis, so for any finite valency the bound fraction is an analytic, strictly monotonic function of receptor density with no singularity: \emph{a superselective threshold is a sharp crossover and never a genuine phase transition.} Language in the literature that describes multivalent binding as switch-like or all-or-none should be read as describing a large but finite logarithmic slope, not a discontinuity. On the other hand, real-rootedness implies that the coefficient sequence $\Omega(k)$ is log-concave,
\begin{equation}\label{eq:logconcave}
\Omega(k)^{2}\ \ge\ \Omega(k-1)\,\Omega(k+1),
\end{equation}
so the equilibrium distribution of bond number is log-concave and hence unimodal for every architecture. There is always a single most probable engagement $k^\star$, never two competing ones, which is what justifies characterising a contact by $k^\star$ at all and is the formal reason the landscapes of Part~II have one bound minimum rather than several.

\subsection{Architecture buys affinity relief, not extra sharpness}

\Cref{fig:topology}(e,f) makes a comparison that is easy to get wrong. The kernels of the four architectures differ by six orders of magnitude, and one might expect the achievable selectivity to differ correspondingly. It does not. When the per-bond free energy is adjusted so that each architecture places its threshold at the same receptor density, the maximum selectivity exponents are $5.8$, $5.6$ and $6.0$ for the flexible, commensurate and incommensurate cases respectively: indistinguishable within the precision of any experiment.

What differs is the price of admission. The per-bond free energy required to reach that common threshold is $-6.4\,\kt$ for flexible ligands against mobile receptors, $-8.0\,\kt$ for a commensurate rigid scaffold, $-8.1\,\kt$ for an incommensurate one and $-10.5\,\kt$ once neighbour exclusion is imposed. A poor topology is paid for in per-bond affinity, at a rate of roughly $4\,\kt$ across the full range, which in ligand terms is about two orders of magnitude in dissociation constant. This reframes what geometric optimisation achieves: matching ligand spacing to receptor spacing does not make a construct intrinsically more discriminating, it makes the required chemistry very much easier. That is a more useful statement for a designer than the usual one, and it is consistent with the observation that low-valency rigid constructs can match flexible high-valency ones provided their spacing is right \cite{Bila2022}.

The exclusion case is the exception that proves the rule. Because \cref{eq:omegarods} truncates at $k=\lceil\ell/2\rceil$, no amount of per-bond affinity can recover the missing states, and the selectivity exponent is capped at $3.7$ rather than $6$. Removing states from the kernel is qualitatively different from making them expensive: the first changes what is achievable, the second changes only what it costs.

\section{Two Units, Naively: The Additive Construction}\label{sec:pairwise}

Let us now do the simplest thing that could possibly work, and assume that the free energies add. Specifically, assume that each formed bond contributes the same free energy $\varepsilon_\zeta$ regardless of how many others are formed, and that the repulsive terms depend on separation alone. The free energy of a configuration is then the sum of a bond term, the combinatorial entropy of \cref{eq:topology} and the repulsive toll,
\begin{equation}\label{eq:additive}
G^{(0)}(\mathbf{k},h)=\sum_{\zeta=1}^{T}k_\zeta\,\varepsilon_\zeta(h)\;-\;\kt\ln\Omega(\mathbf{k})\;+\;W_{\rm rep}(h).
\end{equation}
Summing the Boltzmann weights of \cref{eq:additive} over every engagement state gives the contact partition function, and minus $\kt$ times its logarithm gives the unit-to-unit potential of mean force,
\begin{equation}\label{eq:pairPMF}
\Xi(h)=\sum_{\mathbf{k}}\Omega(\mathbf{k})\,
\exp\Bigl[-\beta\textstyle\sum_\zeta k_\zeta\varepsilon_\zeta(h)\Bigr],
\qquad
W_{AB}(h)=W_{\rm rep}(h)-\kt\ln\Xi(h).
\end{equation}

For a single moiety type the sum collapses to a form worth having explicitly. Using the tethered kernel \cref{eq:omegatether} and writing $b=\rho\pi\lambda^{2}e^{\beta|\varepsilon|}$ for the per-ligand availability, the binomial theorem gives
\begin{equation}\label{eq:compact}
\Xi=\sum_{k=0}^{\ell}\binom{\ell}{k}b^{k}=(1+b)^{\ell},
\qquad
-\kt\ln\Xi=-\ell\,\kt\ln(1+b),
\end{equation}
so the attractive free energy grows linearly in valency and only logarithmically in receptor density. Both dependences matter. The linearity in $\ell$ is why valency is such an effective lever. The logarithm is why the response is not simply proportional to density, and why the interesting behaviour appears in the logarithmic slope
\begin{equation}\label{eq:alpha}
\alpha\equiv\frac{\mathrm{d}\ln\theta}{\mathrm{d}\ln\rho}
=(1-\theta)\,\ell\,\frac{b}{1+b},
\end{equation}
where $\theta$ is convenintely derived  using a Langmuir-Hill isotherm where a multivalent unit binds to a mulitvalent surface, defining a fraction of bound units such as

\begin{equation}\label{eq:theta}
\theta=\frac{z\Xi e^{-\beta W_{\rm rep}}}{(1+z\Xi e^{-\beta W_{\rm rep}})}
\end{equation}

where $z$ is the activity of the multivalent unit. Equation~\eqref{eq:alpha} is the definition of superselectivity made operational: a monovalent binder has $\ell=1$ and hence $\alpha\le1$ for every choice of parameters, so its discrimination between two receptor densities can never exceed their ratio, whereas a multivalent construct is bounded instead by $\alpha\le\ell$.

Equation~\eqref{eq:pairPMF} contains the essential drama of the theory. On one side, a repulsive toll that grows smoothly and inexorably as the units approach. On the other, an attractive term amplified combinatorially, because $\Omega(\mathbf{k})$ counts configurations and its logarithm can grow faster than linearly in the number of available partners. A smooth toll against a superlinearly growing payment is a recipe for a threshold, and thresholds are what make recognition decisive rather than graded. Thresholds of exactly this kind have been measured for multivalent polymers, colloids and DNA nanostructures binding receptor-bearing surfaces \cite{Dubacheva2023,Linne2021,Bila2022}, and exploited to build carriers that read receptor density rather than receptor identity \cite{Tian2020,Liu2020}.

We should be candid about the two assumptions behind \cref{eq:additive}, because both are false, and false in the same way. Bond free energies are \emph{not} independent of engagement: moieties on a shared scaffold cannot bind independently, since the first bond localises the whole unit and thereby changes what entropy remains available to the second. And repulsion is \emph{not} a function of $h$ alone: engaging bonds compresses the corona and strains the scaffold, so the toll depends on $\mathbf{k}$ too. Both failures are the same phenomenon wearing different clothes, and it is time to face it.

\section{Why Additivity Fails, and by How Much}\label{sec:manybody}

\subsection{The body-order expansion}

If pair potentials were the whole story, the free energy of $N$ bodies would be the sum over pairs. Define instead exactly what is left over when all lower-order contributions are subtracted. Writing $W^{(n)}$ for the exact potential of mean force of $n$ bodies, obtained from \cref{eq:pmf}, the irreducible potentials follow recursively,
\begin{align}
w^{(2)}(i,j)&=W^{(2)}(i,j),\label{eq:w2}\\
w^{(3)}(i,j,k)&=W^{(3)}(i,j,k)-\!\!\sum_{\rm pairs}\!w^{(2)},\label{eq:w3}\\
w^{(4)}&=W^{(4)}-\!\!\sum_{\rm triples}\!w^{(3)}-\!\!\sum_{\rm pairs}\!w^{(2)},\label{eq:w4}
\end{align}
and by construction the exact total is the body-order expansion
\begin{equation}\label{eq:expansion}
W^{(N)}=\sum_{i<j}w^{(2)}_{ij}+\sum_{i<j<k}w^{(3)}_{ijk}+\sum_{i<j<k<l}w^{(4)}_{ijkl}+\cdots
\end{equation}
Pairwise additivity is the claim that $w^{(n)}=0$ for all $n\ge3$, and by the third consequence of \cref{eq:pmf} this claim is generically false for a potential of mean force in a condensed medium. The useful questions are how large $w^{(3)}$ is and what sign it takes.

This is not a new worry. The cohesive energy of solid argon needs the three-body Axilrod, Teller and Muto term to come out right \cite{AxilrodTeller1943,Muto1943}, and inclusions in a fluid membrane interact through curvature fields that refuse to superpose once there are three of them \cite{Goulian1993,Reynwar2007,Azadbakht2024}. What is new is that we can pin the effect down exactly in a case that matters for biology.

\subsection{An exactly solvable case: crowding}

Biological media are crowded, and crowding is the one many-body problem in this business that can be solved on paper. The mechanism is Asakura and Oosawa depletion \cite{Asakura1954,Vrij1976,Lekkerkerker2011}: crowders of radius $r_d$ cannot approach a unit of radius $R$ closer than their own radius, so each unit carries an invisible exclusion sphere of radius $a=R+r_d$; when two exclusion spheres overlap, volume is handed back to the crowders, their entropy rises, and the units are pushed together by the osmotic pressure $\Pi=\rho_d\kt$. It is a purely entropic attraction, mediated entirely by the degrees of freedom we integrated out in \cref{eq:pmf}.
The free energy is $-\Pi$ times the volume returned to the crowders, and that volume follows from inclusion and exclusion on the union of the exclusion spheres. For two units the freed volume is the lens of intersection,
\begin{equation}\label{eq:lens}
V_{ij}=\frac{\pi}{12}(4a+d)(2a-d)^2,\qquad d<2a,
\end{equation}
whereas for three units the union is $\sum_iV_i-\sum_{i<j}V_{ij}+V_{123}$, so the freed volume is $\sum_{i<j}V_{ij}-V_{123}$ and
\begin{equation}\label{eq:AO3}
W^{(3)}=-\Pi\Bigl(\sum_{i<j}V_{ij}-V_{123}\Bigr)
=\underbrace{\sum_{i<j}w^{(2)}_{ij}}_{\text{pairwise}}\;+\;\underbrace{\Pi\,V_{123}}_{\textstyle w^{(3)}\;>\;0}.
\end{equation}
Three exact conclusions follow, and each is worth a sentence in plain language.

\begin{figure}[t!]
\centering
\includegraphics[width=\textwidth]{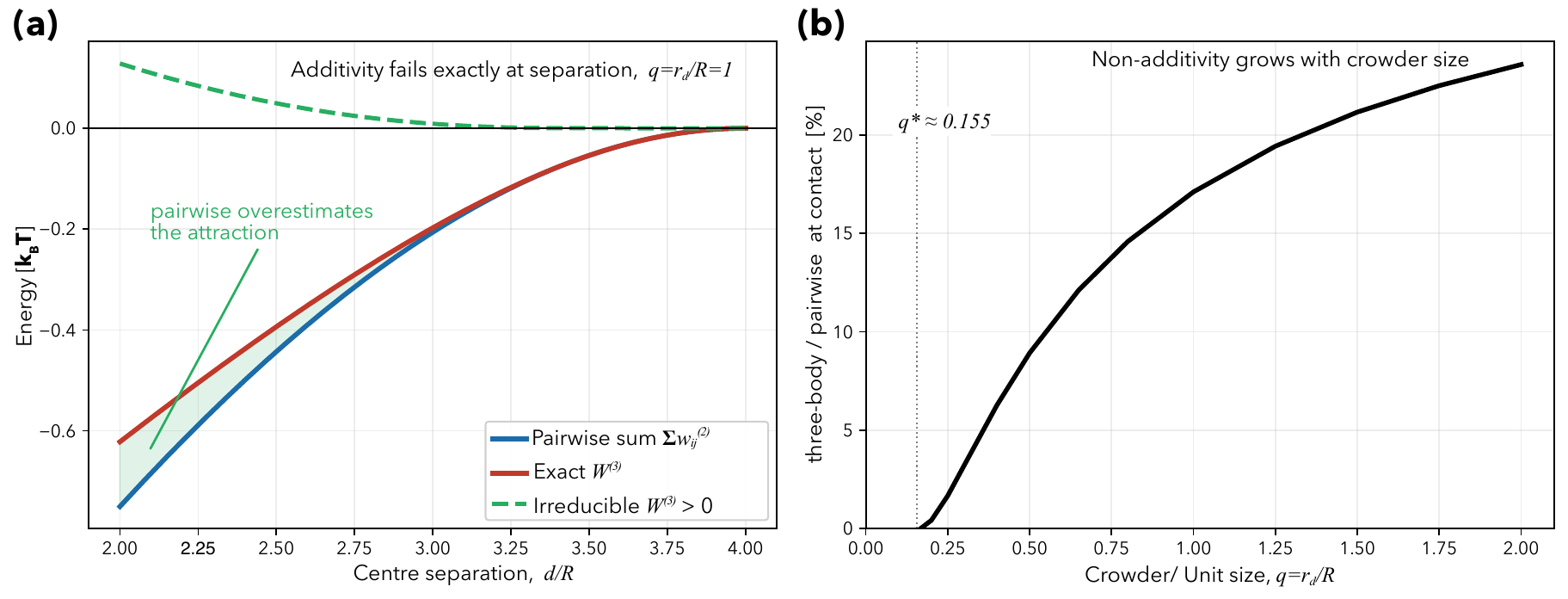}
\caption{Pairwise additivity failing, exactly. \textbf{(a)}~Three units in crowders with size ratio $q=1$: the pairwise sum overestimates the attraction relative to the exact three-body result, the difference being the irreducible term $w^{(3)}=+\Pi V_{123}$ of \cref{eq:AO3}, which is repulsive. \textbf{(b)}~Non-additivity at contact versus crowder size, showing the geometric onset at $q^*\simeq0.155$ of \cref{eq:qstar} and growth to about $24\%$.}
\label{fig:manybody}
\end{figure}

The sign is the surprise. Since $V_{123}\ge0$ we have $w^{(3)}=+\Pi V_{123}\ge0$, so the irreducible three-body depletion interaction \emph{opposes} the pairwise attraction and pairwise treatments of crowding systematically \emph{overestimate} it. The reason is transparent once seen: when three exclusion shells all overlap in a common lens, the pairwise sum counts that shared region three times, whereas only one region was actually freed. The three-body term is the correction for double-counting a benefit, and many-body physics is therefore not automatically cooperative.
The onset is pure geometry. Three exclusion spheres share a common volume at mutual contact only if $a\ge d/\sqrt3$, which for $d=2R$ gives a sharp threshold in the size ratio $q=r_d/R$,
\begin{equation}\label{eq:qstar}
q^*=\frac{2}{\sqrt3}-1\simeq0.155 .
\end{equation}
Crowders smaller than about $15\%$ of the unit radius produce depletion that really is additive, because three shells never manage to share a common volume; larger crowders do not. Since cells and plasma contain crowders spanning $1$ to $20$~nm while the units of interest span $5$ to $100$~nm, biological media sit astride this threshold, and which side a given encounter falls on is decided by a ratio of radii.
The magnitude is not small. The triple-overlap volume $V_{123}$ follows in closed form by slicing the intersection parallel to the plane of the three centres and integrating the resulting circular-triangle area; the expression is validated in two analytic limits and against Monte Carlo, and is implemented in the released code, so the expansion is exact. \Cref{fig:manybody} and \cref{tab:nonadd} give the numbers: at $q=1$ the irreducible three-body term is $17\%$ of the pairwise sum at contact, rising towards $24\%$ for larger crowders. A $17\%$ error in a free energy is not a $17\%$ error in an observable, because free energies enter exponentially. It is a systematic bias in every binding constant computed from a pairwise model.

\begin{table}[t!]
	\centering
	\small
	\caption{Irreducible three-body depletion at mutual contact, as a percentage of the pairwise sum, computed exactly from \cref{eq:AO3}.}
	\label{tab:nonadd}
	\begin{tabular}{@{}lccccccc@{}}
		\toprule
		$q=r_d/R$ & $0.17$ & $0.25$ & $0.4$ & $0.5$ & $1.0$ & $1.5$ & $2.0$\\
		\midrule
		$|w^{(3)}|/|\sum w^{(2)}|$ & $0.0\%$ & $1.7\%$ & $6.3\%$ & $8.9\%$ & $17.1\%$ & $21.2\%$ & $23.6\%$\\
		\bottomrule
	\end{tabular}
\end{table}

\subsection{The full catalogue, and why signs differ}

Depletion is the cleanest case but far from the only one. \Cref{tab:mechanisms} catalogues the mechanisms that generate irreducible many-body terms between the moieties of two associating units, together with the sign each contributes. The heterogeneity of that sign column is the message worth carrying forward: many-body physics does not merely renormalise the pair interaction into a slightly different pair interaction, but can \emph{reverse} the direction of cooperativity depending on mechanism.
Two entries are specific to multivalency and survive even in a perfect solvent, so they deserve emphasis. \emph{Scaffold connectivity} is simply unavoidable: tie several moieties to one body and they stop being independent, because binding one confiscates translational and rotational freedom the others were exploiting. \emph{Corona compression} is its mirror image on the repulsive side, and it is the mechanism by which engagement becomes self-limiting. Notice what that implies. If the cost of the $(k{+}1)$-th bond rises with $k$, then beyond some engagement the next bond is not worth forming, and recognition acquires a \emph{ceiling} as well as a floor. Here, in the foundations, is the seed of upper thresholds, and we have arrived at it without once specifying what the units are for.

\begin{table}[t!]
\centering
\small
\caption{Where irreducible many-body terms come from. Sign refers to $w^{(3)}$ relative to the pairwise attraction: $+$ opposes it (anti-cooperative), $-$ reinforces it (cooperative).}
\label{tab:mechanisms}
\setlength{\tabcolsep}{4pt}
\begin{tabularx}{\textwidth}{@{}>{\raggedright\arraybackslash}p{3.5cm} c L@{}}
\toprule
\textbf{Mechanism} & \textbf{Sign} & \textbf{Physical content}\\
\midrule
Crowder depletion & $+$ & Triple shell overlap is triple-counted by pairs; exact, \cref{eq:AO3}. Quantified for hard-sphere mixtures in ref.~\cite{PerezAngel2025}.\\
Scaffold connectivity & $+$ & Moieties share a body: the first bond localises the unit, reducing the configurational entropy left for the rest. Intrinsic to multivalency.\\
Corona compression & $+$ & Engagement compresses the corona, so the cost of each extra bond grows with the number already formed. Makes engagement self-limiting.\\
Membrane curvature & $\pm$ & Curvature fields of three or more inclusions do not superpose; measured to grow superlinearly and change sign with cluster size \cite{Goulian1993,Reynwar2007,Weikl2018,Azadbakht2024}.\\
Nonlinear screening & $\pm$ & Poisson and Boltzmann is nonlinear, so double layers do not add.\\
Hydration structure & $\pm$ & Water layering between three surfaces is not the sum of pair layerings.\\
Partner recruitment & $-$ & Binding locally concentrates mobile partners, raising the density that later bonds experience. Cooperative.\\
Lateral partner attraction & $-$ & Weak attraction between partners, even below $\kt$, sharpens thresholds almost step-wise \cite{Curk2025}.\\
Allostery & $\pm$ & Binding alters conformation, hence the affinity of the remaining sites.\\
\bottomrule
\end{tabularx}
\end{table}

\subsection{Putting the correction to work}

We do not need the full expansion to make progress. Collecting the mechanisms of \cref{tab:mechanisms} into a symmetric coupling matrix $\mathbf{J}$ and truncating at the leading non-additive order, the free energy of a configuration becomes
\begin{equation}\label{eq:withMB}
G(\mathbf{k},h)=G^{(0)}(\mathbf{k},h)\;+\;\sum_{\zeta\le\zeta'}J_{\zeta\zeta'}\,k_\zeta k_{\zeta'}\;+\;\mathcal{O}(k^3),
\end{equation}
with $J_{\zeta\zeta'}<0$ where cooperative mechanisms dominate, meaning recruitment and lateral attraction, and $J_{\zeta\zeta'}>0$ where anti-cooperative ones do, meaning depletion, connectivity and corona compression. The diagonal $J_{\zeta\zeta}>0$ is what makes engagement self-limiting.
It matters that $\mathbf{J}$ is not a fitting device. Every element has an identified microscopic origin in \cref{tab:mechanisms} and can be estimated independently: by geometry, as we did exactly for depletion in \cref{eq:AO3}; by elasticity for the corona and membrane terms; or by coarse-grained and machine-learned effective potentials \cite{CGnano2023,Drautz2019}. Where the couplings are switched off, \cref{eq:withMB} reduces to \cref{eq:additive} and the standard theory is recovered.

One last observation, offered because it is genuinely useful rather than merely decorative. An interaction among $n$ moieties is not a collection of links; it is a single object joining $n$ things at once. Mathematics has a word for that, a hyperedge, and an entire apparatus for handling networks built from them \cite{Battiston2021}. The many-body problem of molecular association and the higher-order-interaction problem of complex systems are, formally, the same problem in different notation. That is a good sign, because it means the tools developed for one can be borrowed for the other, just as the matching-polynomial results of \cref{sec:heilmann} were borrowed from the theory of monomer--dimer systems.

\section{The General Functional}\label{sec:functional}

We can now write down what Part~I was built to produce. Assembling the obligatory repulsion of \cref{sec:necessity}, the topology kernel of \cref{sec:topology}, the specific bond free energies carrying the contrast of \cref{eq:specificity}, and the many-body couplings of \cref{eq:withMB}, the association free energy of two multivalent units with descriptor sets $\boldsymbol{\Lambda}_A,\boldsymbol{\Lambda}_B$ in the medium of \cref{tab:medium} is
\begin{equation}\label{eq:master}
W_{AB}(h)=
\underbrace{W_{\rm rep}(h)}_{\substack{\text{non-specific}\\ \text{sets the threshold}}}
\;-\;\kt\ln\!\!\underbrace{\sum_{\mathbf{k}}\Omega(\mathbf{k};\mathbf{g},\xi)\,
\exp\Bigl[-\beta\Bigl(\sum_\zeta k_\zeta\varepsilon_\zeta(h)
+\!\!\sum_{\zeta\le\zeta'}\!J_{\zeta\zeta'}k_\zeta k_{\zeta'}\Bigr)\Bigr]}_{\substack{\text{specific, combinatorially amplified,}\\ \text{with irreducible many-body coupling}}}
\end{equation}
Every term in \cref{eq:master} is separately measurable and separately designable, and three limits organise what follows.

In the \emph{monovalent} limit $\ell=1$ the kernel is trivial, $\mathbf{J}$ is irrelevant, and \cref{eq:master} yields a single sigmoidal binding curve with $\alpha\le1$: classical affinity, and no threshold. In the \emph{additive multivalent} limit $\mathbf{J}=0$ it reduces to \cref{eq:pairPMF} and recovers the standard combinatorial-entropy theory of multivalent binding \cite{Kitov2003,Martinez-Veracoechea2011}, with thresholds driven by $\ln\Omega$ alone and their sharpness bounded by the matching-polynomial result of \cref{sec:heilmann}. In the \emph{full many-body} regime thresholds sharpen or blunt according to the sign of $\mathbf{J}$, and because connectivity and corona compression contribute $J>0$ growing with $k$, engagement can become self-limiting and bound recognition from above as well as below.

\begin{figure}[t!]
\centering
\includegraphics[width=\textwidth]{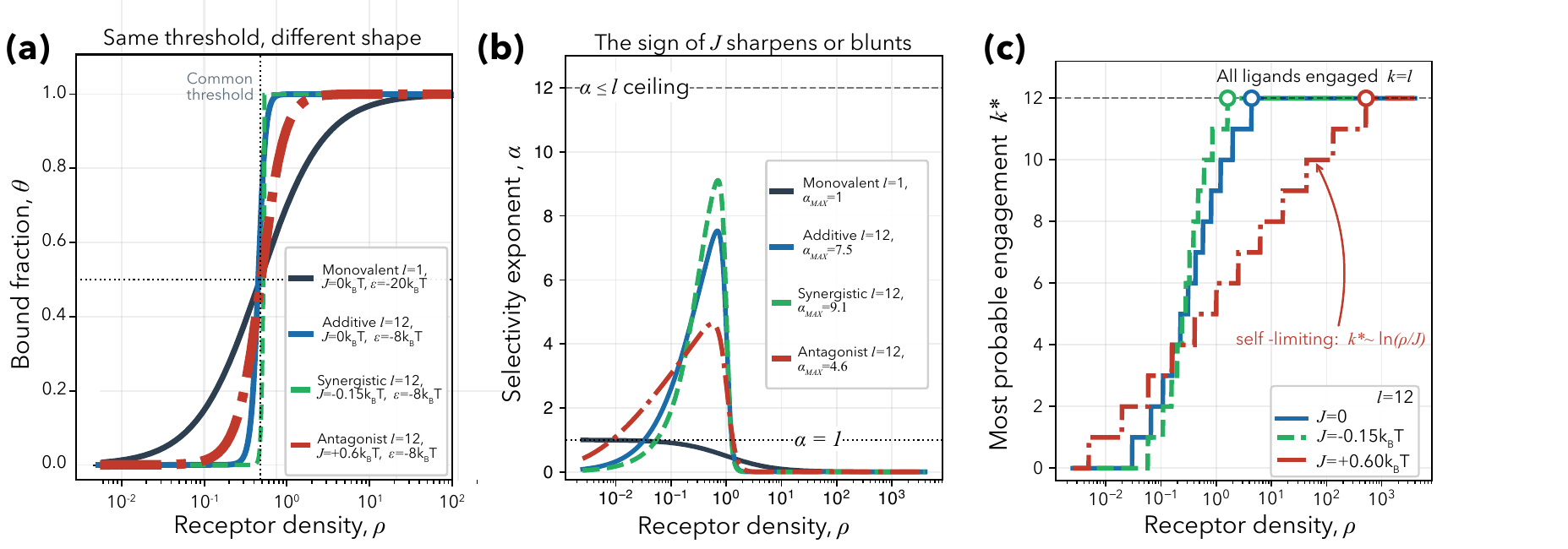}
\caption{The three limits of the general functional, \cref{eq:master}. \textbf{(a--c)}~Schematics: a monovalent unit, for which the topology kernel is trivial and the coupling matrix has nothing to couple; an additive multivalent unit, whose bonds ignore one another; and a many-body unit, in which each bond formed alters the free energy of the next. \textbf{(d)}~Bound fraction against receptor availability. The per-bond free energy $\varepsilon$ of each case is solved so that all four cross $\theta=1/2$ together, and is quoted in the legend: monovalency is expensive. \textbf{(e)}~Selectivity exponent. The monovalent curve saturates the classical bound $\alpha\le1$ exactly; attractive coupling sharpens beyond the additive result, repulsive coupling blunts it. \textbf{(f)}~Most probable engagement. With $\mathbf{J}=0$, $k^\star$ reaches full valency as soon as receptors are plentiful; with $J>0$ it grows only as $\ln\rho/J$, so engagement is rationed and recognition is bounded from above. Kernel $\Omega(k)=\binom{\ell}{k}$ with $\ell=12$, corona toll $W_{\rm rep}=9\,\kt$, fugacity $z=10^{-3}$; all partition sums evaluated in log space.}
\label{fig:regimes}
\end{figure}

\Cref{fig:regimes} makes the three limits concrete for a single receptor type against the same corona toll $W_{\rm rep}=9\,\kt$, with the per-bond free energy in each case solved so that all four curves cross $\theta=1/2$ at the same receptor availability. Comparing them at a common threshold is the only fair comparison, because otherwise a difference in position masquerades as a difference in kind. Three things are then visible at once.

The first is the price of monovalency. A single binding group must supply $\varepsilon=-22.8\,\kt$ to clear a toll that twelve cooperating groups clear with $-7.9\,\kt$ each, a difference of some six orders of magnitude in dissociation constant. This is the quantitative content of the remark in \cref{sec:necessity} that repulsion is what makes multivalency worth having: without a toll to pay there would be nothing for the combinatorial entropy to buy. The second is that the monovalent curve has $\alpha_{\max}=1.00$ exactly, as it must, while the additive twelve-valent construct reaches $7.53$. Affinity and selectivity are different quantities, and only the second is architectural.

The third is the effect of the coupling, and it is asymmetric in a way worth noting. A modest attractive coupling, $J=-0.15\,\kt$, sharpens the threshold from $\alpha_{\max}=7.53$ to $9.11$ and lets the same threshold be reached with slightly weaker bonds, because each bond formed subsidises the next. A repulsive coupling of $J=+0.60\,\kt$ does the reverse twice over: it blunts the threshold to $\alpha_{\max}=4.63$ and demands stronger bonds, $-9.7\,\kt$, to reach the same midpoint. More interesting is what it does to the engagement itself. With $\mathbf{J}=0$ the optimal engagement $k^\star$ climbs to full valency as soon as receptors are plentiful, whereas with $J>0$ the marginal cost of the next bond grows as $Jk$ and the optimum instead creeps up logarithmically, $k^\star\simeq\ln\rho/J$, so that reaching full valency requires roughly $120$ times more receptors. Engagement is rationed rather than switched on. That is the precise sense in which many-body coupling bounds recognition from above as well as below, and it is a design lever rather than a defect: a construct whose corona stiffens as it engages cannot commit itself irreversibly to the first surface it meets.

That is the general theory of association at equilibrium. It is worth pausing on how little it assumes: two objects, some binding groups, a repulsive layer, and water. Nothing about biology, nothing about medicine, nothing about what the units are made of. What it cannot yet do is tell us how long anything takes, and that turns out to matter enormously, because several of the most important discriminations biology performs are not thermodynamic at all. That is the business of Part~II.

\part*{\centering Part II \quad Time: Dynamics and the Non-Equilibrium Nature of Recognition}
\addcontentsline{toc}{section}{\textbf{Part II. Time: Dynamics and the Non-Equilibrium Nature of Recognition}}

\section{Why Equilibrium Is Not Enough}\label{sec:whytime}

Part~I produced a free energy, and a free energy answers exactly one question: given unlimited time, what fraction of encounters end up bound? For a great deal of chemistry that is the right question. For biology it very often is not, and there are three independent reasons.

The first is that biology does not have unlimited time. A nanoparticle in the bloodstream passes a given endothelial surface in a fraction of a second. A T cell scans a dendritic cell for minutes, examining thousands of peptides. A virion has one attempt before mucociliary clearance removes it. If the encounter is shorter than the time needed to reach equilibrium, then what matters is not the depth of the free-energy minimum but the rate of descent into it. Two constructs with identical equilibrium constants can behave completely differently under flow, and the one that wins is the one that engages faster.

The second reason is that a deep minimum is not a bound state if the system cannot get into it, and not a useful bound state if it cannot get out. Part~I already showed that a bare surface in physiological salt falls into a well $21\,\kt$ deep. Thermodynamically that is spectacular binding. Functionally it is a disaster, because nothing that binds that hard can ever be released, recycled or regulated. Living systems need contacts that last a specified time, and the specified time is a kinetic quantity that the equilibrium constant does not determine.

The third reason is the deepest. Some biological discriminations are provably impossible at equilibrium and are achieved only by spending energy. The classic example, which we take up in Part~III, is T cell antigen recognition, where discrimination between a pathogenic peptide and a self peptide differing by a single residue far exceeds what their binding free energies would permit through \cref{eq:discrim}. The trick is to make the decision depend on how long a complex survives rather than on how favourable it is, and to read that lifetime out through a sequence of energy-consuming steps. This is kinetic proofreading, and it is a non-equilibrium mechanism in the strict sense: switch off the energy supply and the discrimination vanishes.

So we need to put time into the theory. Fortunately Part~I already told us what kind of dynamics to expect. \Cref{tab:medium} showed that motion at this scale is overdamped, with a Reynolds number so small that inertia never matters, and that the solvent relaxes thousands of times faster than the units move. Those two facts together specify the equation of motion almost uniquely.

\section{From Master Equation to Smoluchowski}\label{sec:smol}

There are two natural coordinates, and they call for two different but equivalent descriptions. Engagement $\mathbf{k}$ is a discrete integer count, so its evolution is naturally a jump process. Separation $h$ is continuous, so its evolution is naturally a diffusion in a potential.

Let $P(\mathbf{k},t)$ be the probability of engagement $\mathbf{k}$ at time $t$. Conservation of probability over the discrete state space gives the master equation
\begin{equation}\label{eq:mastereq}
\frac{\partial P(\mathbf{k},t)}{\partial t}
=\sum_{\mathbf{k}'}\bigl[\,W(\mathbf{k}'\!\to\!\mathbf{k})P(\mathbf{k}',t)-W(\mathbf{k}\!\to\!\mathbf{k}')P(\mathbf{k},t)\bigr],
\end{equation}
and the rates are not free to choose. Any closed system relaxing to the equilibrium of \cref{eq:withMB} must satisfy detailed balance with respect to it,
\begin{equation}\label{eq:detbal}
\frac{W(\mathbf{k}\!\to\!\mathbf{k}')}{W(\mathbf{k}'\!\to\!\mathbf{k})}
=\exp\!\bigl[-\beta\bigl(\mathcal{G}(\mathbf{k}')-\mathcal{G}(\mathbf{k})\bigr)\bigr].
\end{equation}
For single-bond formation and rupture a convenient parametrisation consistent with \cref{eq:detbal} is
\begin{equation}\label{eq:rates}
W(k\!\to\!k{+}1)=k_{\rm on}\,(R-k)\,\frac{\Omega(k{+}1)}{\Omega(k)}\,e^{-\beta\Delta\mathcal{G}^{\ddagger}_{\rm on}},
\qquad
W(k\!\to\!k{-}1)=k_{\rm off}\,k ,
\end{equation}
in which the topology kernel appears explicitly as a ratio, so that architecture enters the kinetics as well as the thermodynamics. Equation~\eqref{eq:detbal} is a constraint rather than a modelling choice, and rates that violate it describe a system driven by an external energy source. That is precisely what kinetic proofreading requires, and we return to it in \cref{sec:proofreading}.

For the continuous coordinate the overdamped limit of the Fokker and Planck equation is the Smoluchowski equation \cite{Smoluchowski1917,Risken1989},
\begin{equation}\label{eq:smol}
\frac{\partial P(h,t)}{\partial t}
=\frac{\partial}{\partial h}\!\left[D(h)\left(\frac{\partial P}{\partial h}+\beta P\,\frac{\partial W_{AB}}{\partial h}\right)\right]
\equiv-\frac{\partial \mathcal{J}}{\partial h},
\end{equation}
with $W_{AB}$ the potential of mean force of \cref{eq:master}, $D(h)$ the position-dependent relative diffusivity and $\mathcal{J}$ the probability flux. Equation~\eqref{eq:smol} may be rewritten in a manifestly detailed-balanced form that makes its equilibrium transparent,
\begin{equation}\label{eq:smol2}
\frac{\partial P}{\partial t}=\frac{\partial}{\partial h}\left[D\,e^{-\beta W_{AB}}\frac{\partial}{\partial h}\Bigl(e^{\beta W_{AB}}P\Bigr)\right],
\end{equation}
from which the zero-flux stationary solution is read off immediately as $P_{\rm eq}(h)\propto e^{-\beta W_{AB}(h)}$. Equilibrium statistical mechanics is therefore recovered as the long-time limit, and everything in Part~I is contained in Part~II rather than replaced by it. A stationary state with $\mathcal{J}\neq0$, by contrast, is a non-equilibrium steady state sustained only by driving.

The quantity we actually want is a timescale. Integrating \cref{eq:smol} twice with an absorbing boundary gives the mean first-passage time from $h_0$ to a target at $h_t$,
\begin{equation}\label{eq:mfpt}
\tau(h_0\!\to\!h_t)=\int_{h_t}^{h_0}\!\frac{\mathrm{d}y}{D(y)}\,e^{\beta W_{AB}(y)}\!\int_{y}^{\infty}\!\mathrm{d}x\;e^{-\beta W_{AB}(x)},
\end{equation}
and when the landscape has a well of depth $\Delta\mathcal{G}^{\ddagger}$ separated from the outside by a sharp barrier, the double integral is dominated by its endpoints and reduces to the Kramers form \cite{Kramers1940,Hanggi1990}
\begin{equation}\label{eq:kramers}
\tau\simeq\tau_0\,e^{\beta\Delta\mathcal{G}^{\ddagger}},
\qquad \tau_0\sim1\ \text{ns for nanoscale units.}
\end{equation}

Equation~\eqref{eq:kramers} is the single most consequential result in Part~II, because it says that times depend \emph{exponentially} on free energies. A free energy that changes by a factor of two changes a rate by its exponential, and since the free energies in question are tens of $\kt$, the rates span many orders of magnitude. Recognition thresholds that look modest on a thermodynamic plot become ferociously sharp when read as lifetimes.

\section{Kinetic Selectivity: Reading Density Through Lifetime}\label{sec:kinetic}

\begin{figure}[t!]
\centering
\includegraphics[width=\textwidth]{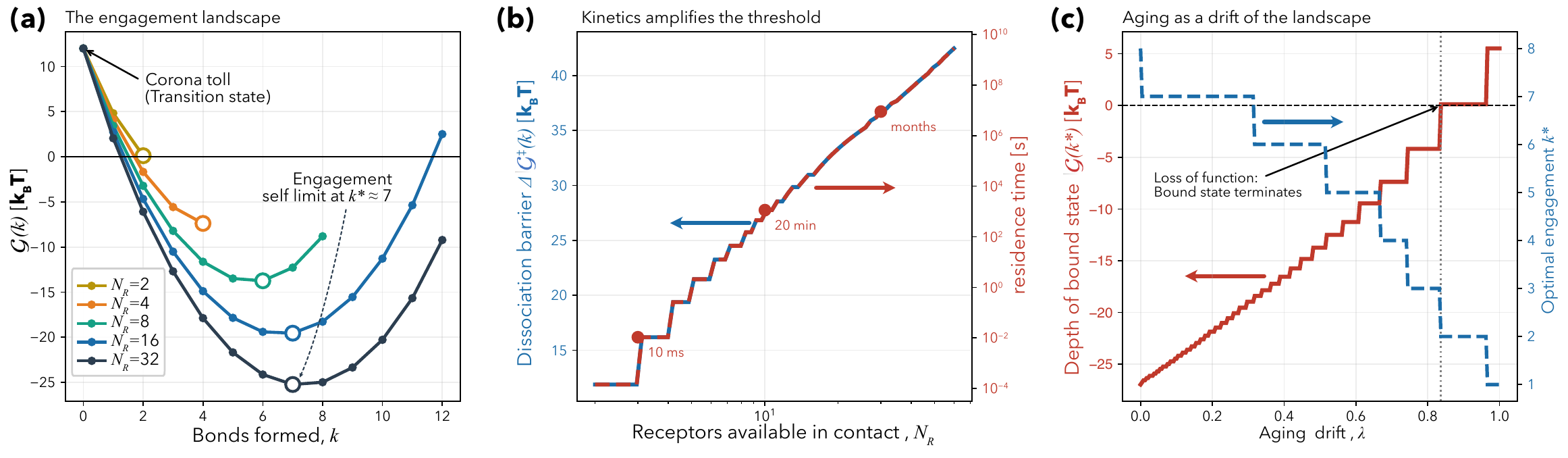}
\caption{Dynamics of a multivalent contact, computed from \cref{eq:withMB} with $\ell=12$, $\varepsilon=-4\,\kt$, corona toll $W_{\rm rep}=12\,\kt$ and self-coupling $J=1\,\kt$. \textbf{(a)}~The engagement landscape. Approach must first pay the corona toll at $k=0$, which acts as the transition state; the bound minimum then deepens with the number of available receptors $R$. Because $J>0$, engagement self-limits at $k^\star\approx7$ no matter how many receptors are offered. \textbf{(b)}~Dissociation barrier and the resulting residence time from \cref{eq:kramers}. A tenfold change in receptor number moves the lifetime of the contact from ten milliseconds to months. \textbf{(c)}~Ageing as a slow drift. As receptors are lost the bound minimum rises, and at a critical drift it crosses zero and disappears in the saddle-node bifurcation of \cref{eq:bifurc}.}
\label{fig:dynamics}
\end{figure}

\Cref{fig:dynamics}(a) shows the engagement landscape for the construct of Part~I, and its shape is worth reading carefully because every feature was predicted rather than assumed. At $k=0$ the free energy sits at the full corona toll, because the units are in contact but no bond has yet formed: the repulsive layer that \cref{sec:necessity} showed to be obligatory now doubles as the \emph{transition state} of the association reaction. Bonds then form, each paying for itself and for the combinatorial entropy it unlocks, until the anti-cooperative coupling $J>0$ of \cref{eq:withMB} makes further bonds unprofitable. The minimum sits at $k^\star\approx7$ and, strikingly, stays there as the receptor supply is increased eightfold. This is the ceiling on engagement that \cref{tab:mechanisms} predicted from corona compression and scaffold connectivity, and it appears here without any additional assumption. That the minimum is unique rather than one of several is guaranteed by the log-concavity of \cref{eq:logconcave}.
\Cref{fig:dynamics}(b) converts landscape depth into time using \cref{eq:kramers}, and the numbers are startling. With three receptors available the contact lives about ten milliseconds. With ten it lives twenty minutes. With thirty it lives months. A factor of ten in receptor number spans nine orders of magnitude in residence time. Nothing in the thermodynamics changes character across that range; the free energy simply grows smoothly. It is the exponential in \cref{eq:kramers} that turns a smooth thermodynamic gradient into a switch.
This has a direct practical consequence, and it is one of the more useful results in this work. If a construct encounters a surface only briefly, as under flow, then the relevant question is whether it can form enough bonds during the encounter to survive it. Since the number of bonds achievable in a fixed time depends on receptor density through the rates of \cref{eq:rates}, and the survival probability depends exponentially on the number of bonds, selectivity measured kinetically can be considerably sharper than selectivity measured at equilibrium. Experiments on multivalent polymers have found exactly this, with initial association more strongly density-selective than the final equilibrium state \cite{Ravnik2026}. Design for a brief encounter and design for a long incubation are therefore different problems with different optima.

\section{Breaking Detailed Balance: Kinetic Proofreading}\label{sec:proofreading}

Everything so far respects \cref{eq:detbal}, and everything so far is therefore limited by a hard ceiling: at equilibrium the ratio of bound populations for two competing ligands can never exceed the ratio of their Boltzmann factors, which is the content of \cref{eq:discrim}. If two peptides differ by $2\,\kt$ in binding free energy, no equilibrium mechanism, however elaborate, can discriminate them better than sevenfold. Biology routinely does much better, and the resolution is that biology does not play by equilibrium rules.

The mechanism is due to Hopfield and Ninio and was adapted to immune recognition by McKeithan \cite{Hopfield1974,Ninio1975,McKeithan1995,Francois2013}. Insert a series of $N$ irreversible, energy-consuming modification steps between binding and response, each proceeding at rate $k_p$, and suppose the complex is lost at rate $k_{\rm off}=\tau^{-1}$ at every stage. The probability of surviving one step is the competition between progression and loss, $\frac{k_p}{k_p+k_{\rm off}}$, and since the steps are independent the probability of completing all $N$ is
\begin{equation}\label{eq:proof}
P_{\rm signal}(\tau)=\left(\frac{k_p\tau}{1+k_p\tau}\right)^{\!N}.
\end{equation}
The discrimination between two ligands with lifetimes $\tau_1>\tau_2$ follows by division,
\begin{equation}\label{eq:proofdisc}
\frac{P_{\rm signal}(\tau_1)}{P_{\rm signal}(\tau_2)}
=\left(\frac{\tau_1}{\tau_2}\cdot\frac{1+k_p\tau_2}{1+k_p\tau_1}\right)^{\!N}
\;\xrightarrow[k_p\tau\ll1]{}\;\left(\frac{\tau_1}{\tau_2}\right)^{\!N},
\end{equation}
a \emph{power law} in the lifetime ratio whose exponent is the number of proofreading steps. The numbers are worth spelling out because they explain something otherwise mysterious. A tenfold difference in complex lifetime corresponds to only about $2.3\,\kt$ of binding free energy and hence to at most tenfold discrimination at equilibrium, but becomes $10^N$ with proofreading: three steps give a thousandfold, five give a hundred thousandfold.

Two caveats belong with \cref{eq:proof}. The mechanism requires the modification steps to be irreversible, so it violates \cref{eq:detbal} and consumes free energy; it is unavailable at equilibrium, and switching off the energy supply abolishes the discrimination. And there is a cost in sensitivity, since $P_{\rm signal}\le1$ falls with $N$: specificity is bought with signal amplitude, and a cell that proofreads more stringently needs more antigen to respond at all. This trade-off is unavoidable, and Part~III shows how immune memory resolves it.

It is worth being clear about what has and has not changed. The free-energy landscape of Part~I is still there, and still governs the lifetime $\tau$ through \cref{eq:kramers}. What proofreading adds is a \emph{readout} exquisitely sensitive to that lifetime, purchased with metabolic energy. The thermodynamics sets the raw signal, the non-equilibrium machinery sets the gain.

\section{Ageing: The Landscape Itself Drifts}\label{sec:ageing}

There is one more timescale, far slower than any considered so far, and it changes the questions we can ask. Over months and years the parameters in \cref{eq:master} are not constant. Receptor expression falls or rises, the glycocalyx thickens, membranes stiffen as their lipid composition shifts, proteins oxidise and the extracellular matrix cross-links. Collect these slow variables into a vector $\boldsymbol{\lambda}(t)$, so that $\mathcal{G}=\mathcal{G}(\mathbf{k},h;\boldsymbol{\lambda})$, and the joint probability density evolves under two operators acting on widely separated timescales,
\begin{equation}\label{eq:agingFP}
\frac{\partial P(\mathbf{k},h,\boldsymbol{\lambda},t)}{\partial t}
=\underbrace{\mathcal{L}_{\rm fast}\,P}_{\text{\cref{eq:mastereq,eq:smol} at fixed }\boldsymbol{\lambda}}
\;+\;\underbrace{\mathcal{L}_{\boldsymbol{\lambda}}\,P}_{\text{slow drift}},
\end{equation}
where the slow operator is itself of Fokker and Planck form,
\begin{equation}\label{eq:Llam}
\mathcal{L}_{\boldsymbol{\lambda}}P=-\sum_\nu\frac{\partial}{\partial\lambda_\nu}\bigl[V_\nu(\boldsymbol{\lambda})P\bigr]
+\sum_{\nu\mu}\frac{\partial^2}{\partial\lambda_\nu\partial\lambda_\mu}\bigl[D^{\lambda}_{\nu\mu}P\bigr],
\end{equation}
with $V_\nu$ the deterministic drift velocity of each control parameter and $D^\lambda$ its stochastic part \cite{Gardiner2009}. Because $|V|$ is small compared with every binding rate, the fast variables remain quasi-equilibrated on the slowly deforming landscape, which is an adiabatic approximation and the reason the two operators can be treated in sequence rather than together.

The interesting behaviour is not the drift but its endpoint. The bound state persists while $\mathcal{G}(k^\star;\boldsymbol{\lambda})<0$, and it is lost when the minimum and the barrier merge and annihilate,
\begin{equation}\label{eq:bifurc}
\mathcal{G}(k^\star;\boldsymbol{\lambda}_c)=0
\quad\text{and}\quad
\frac{\partial^2\mathcal{G}}{\partial k^2}\Big|_{k^\star}\to0,
\end{equation}
which is a saddle-node bifurcation. \Cref{fig:dynamics}(c) shows this happening. As the drift parameter increases and receptors are progressively lost, the bound state becomes steadily shallower; for most of the trajectory nothing dramatic occurs, and the contact simply weakens and its lifetime shortens through \cref{eq:kramers}. Then, at a critical drift, the bound minimum crosses zero and ceases to exist. Recognition does not fade away, it switches off.

This is worth dwelling on because it reframes a familiar clinical observation. Physiological decline is usually gradual, and then something fails suddenly. The framework here says that both statements can be true simultaneously and without contradiction: the \emph{parameters} drift smoothly while the \emph{state} changes discontinuously, because the map from parameters to state contains a fold. Systems poised near such a fold are also, necessarily, hypersensitive, since the curvature of the landscape vanishes there and fluctuations grow. Critical slowing down of this kind is a generic early-warning signature of an approaching tipping point and is in principle measurable \cite{Scheffer2009}.

\section{What Part II Adds}

Adding time changes four things. Recognition thresholds acquire exponential sharpness through \cref{eq:kramers}, so that a tenfold change in receptor number can span nine orders of magnitude in contact lifetime. Selectivity becomes encounter-dependent, and constructs optimised for brief encounters under flow differ from those optimised for long incubation. Discrimination beyond the equilibrium ceiling of \cref{eq:discrim} becomes possible, but only by breaking detailed balance and paying for it with metabolic energy. And the landscape itself becomes a slow dynamical variable, so that graceful parameter drift produces abrupt functional failure at the bifurcation of \cref{eq:bifurc}. With these in hand we can turn to biology, which uses all four.

\part*{\centering Part III \quad Three Biological Systems}
\addcontentsline{toc}{section}{\textbf{Part III. Three Biological Systems}}

\section{Reading Biology With One Set of Equations}\label{sec:bio-intro}

We now apply the framework to three systems that a biologist would not naturally group together: antibodies, lipoproteins and T cells. They differ in every obvious respect. One is a soluble protein, one a lipid assembly, one a cell. Their sizes span two orders of magnitude and their functions have nothing in common. What they share is architecture. Each is a scaffold presenting several binding moieties to a partner surface, wrapped in something repulsive, negotiating in water. Each is therefore a multivalent unit in the sense of \cref{eq:unit}, and each obeys \cref{eq:master}.
The point of this Part is not to reproduce known immunology or lipid biology. It is to show that three phenomena usually explained by three separate vocabularies, namely antibody avidity, lipoprotein clearance and T cell specificity, are the same physics with different parameters. Where the framework earns its keep is in the numbers: in each case a single computation from \cref{eq:master} or \cref{eq:proof} reproduces a quantitative feature that the standard qualitative account leaves unexplained.
\Cref{tab:bio} places the three systems side by side in the language of \cref{eq:descriptors}. The differences that matter are in three columns: valency, per-bond free energy, and whether the discrimination is thermodynamic or kinetic.

\begin{table}[t!]
\centering
\footnotesize
\caption{Three biological systems as multivalent units. $\ell$ is the effective valency towards the partner surface, $\varepsilon$ the per-bond free energy, and the last column states whether selectivity is achieved at equilibrium or requires driving.}
\label{tab:bio}
\setlength{\tabcolsep}{3.5pt}
\renewcommand{\arraystretch}{1.15}
\begin{tabularx}{\textwidth}{@{}>{\raggedright\arraybackslash}p{1.9cm}
                              >{\raggedright\arraybackslash}p{2.1cm}
                              >{\raggedright\arraybackslash}p{2.3cm}
                              >{\centering\arraybackslash}p{1.0cm}
                              >{\raggedright\arraybackslash}p{1.6cm} L@{}}
\toprule
\textbf{Unit} & \textbf{Scaffold} & \textbf{Moieties} & $\boldsymbol{\ell}$ & $\boldsymbol{\varepsilon}\,[\kt]$ & \textbf{Selectivity}\\
\midrule
IgG, IgE, IgD & Y-shaped Fab arms & antigen-binding sites & $2$ & $-6$ to $-12$ & at equilibrium; avidity from valency\\
IgA dimer & two monomers, J chain & antigen-binding sites & $4$ & $-5$ to $-10$ & at equilibrium; mucosal cross-linking\\
IgA tetramer & four monomers, J chain & antigen-binding sites & $8$ & $-5$ to $-10$ & at equilibrium; needs repetitive antigen\\
IgM pentamer & pentamer, 10 arms & antigen-binding sites & $10$ & $-4$ to $-8$ & at equilibrium; needs repetitive antigen\\
IgM hexamer & hexamer, 12 arms & antigen-binding sites & $12$ & $-4$ to $-8$ & at equilibrium; outperforms at low epitope density\\
LDL & lipid core, ApoB-100 & one receptor-binding region & $1$ & $\approx-20$ & at equilibrium; one strong contact\\
Remnant, VLDL & lipid core, ApoE copies & several ApoE sites & $4$ to $8$ & $\approx-7$ & at equilibrium; avidity from valency\\
T cell & membrane, synapse & TCR against pMHC & $1$ to few & $-6$ to $-10$ & \textbf{kinetic}; proofread, driven\\
\bottomrule
\end{tabularx}
\end{table}

\section{Immunoglobulins: Antigen Size Chooses the Effector}\label{sec:ig}

An antibody is the cleanest multivalent unit in biology, and it is unusual in having two functional faces. The Fab arms engage antigen; the Fc stem engages receptors on effector cells. Both faces are multivalent, both obey \cref{eq:compact}, and the interesting physics lies in how they are coupled.

\subsection{The antigen face}

\begin{figure}[t!]
\centering
\includegraphics[width=\textwidth]{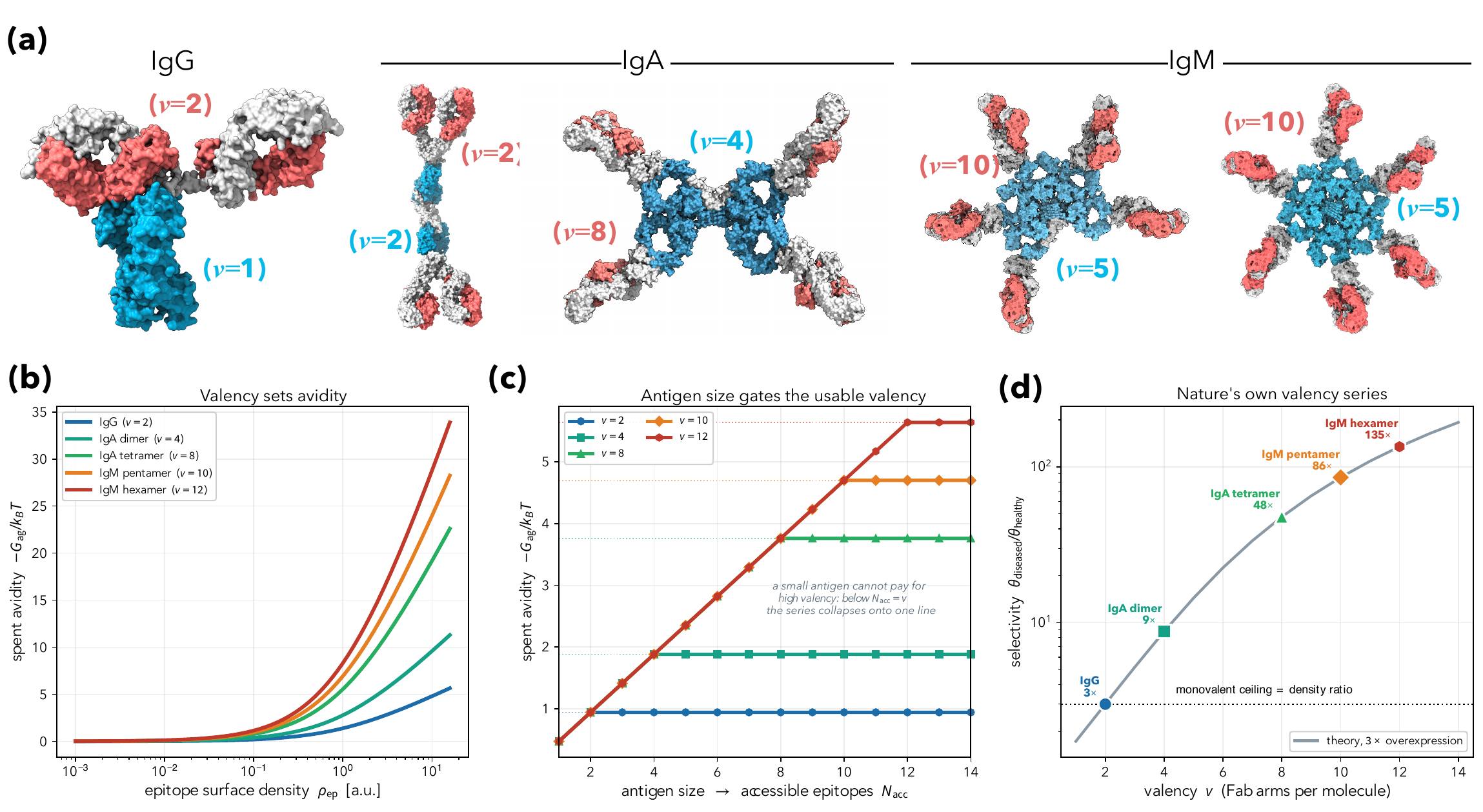}
\caption{\textbf{(a)}~The antigen face of immunoglobulins, from \cref{eq:compact}, across the full isotype valency series: IgG monomer ($\ell=2$), secretory IgA dimer ($\ell=4$) and tetramer ($\ell=8$), IgM pentamer ($\ell=10$) and hexamer ($\ell=12$). Valencies count Fab arms per assembled molecule; IgE and IgD are monomeric and bivalent and so coincide with IgG . \textbf{(b)}~Avidity against epitope density. Valency multiplies avidity. \textbf{(c)}~Antigen size, expressed as the number of accessible epitopes $N_{\rm acc}$, gates how much of that valency can be used: below $N_{\rm acc}=\ell$ the series collapses onto one line and every isotype reverts to its monovalent affinity. \textbf{(d)}~The same series on the selectivity axis of \cref{eq:alpha}, against a threefold difference in epitope density. Bivalent IgG sits exactly on the monovalent ceiling of \cref{eq:discrim}; the IgM hexamer clears it by a factor of $45$ while requiring per-bond free energy weaker by $8\,\kt$.}
\label{fig:ig}
\end{figure}

The immunoglobulin isotypes are, conveniently for us, a valency series rather than an affinity series, and they span it almost uniformly. IgG, IgE and IgD are monomeric and bivalent. Secretory IgA is assembled around the joining chain into a dimer bearing four Fab arms and, less abundantly, into higher polymers; atomic-resolution structures exist for the dimeric and tetrameric forms, the latter bearing eight \cite{Bharathkar2020,Kumar2020}. IgM is secreted as a J-chain-containing pentamer with ten arms \cite{Czajkowsky2009} and, when the joining chain is absent, predominantly as a hexamer with twelve \cite{Oskam2022}. Nature therefore supplies $\ell=2,4,8,10,12$ built from the same monomer, which is as close to a controlled valency titration as biology offers (see structures in \Cref{fig:ig}(a)).

On the antigen side, valency does what \cref{eq:compact} says it should: avidity grows linearly in the number of arms. \Cref{fig:ig}(b) shows the consequence. At the same epitope density the IgM hexamer accumulates six times the avidity of a bivalent IgG, the pentamer five times and the IgA dimer twice, which is why IgM is the first-response antibody, deployed before affinity maturation \cite{Victora2012} has had time to improve any individual site. Valency substitutes for affinity, exactly as the theory requires.
But there is a catch, and it is geometric. Avidity is only realised if the antigen is large enough and repetitive enough to present epitopes to every arm. \Cref{fig:ig}(c) makes the point by capping the usable valency at the number of accessible epitopes $N_{\rm acc}$. Below $N_{\rm acc}=\ell$ the entire series collapses onto one line, so against a small soluble antigen with a single epitope every isotype reverts to the same monovalent affinity and IgM's twelve arms are worthless. The series separates only once the antigen can pay for the valency, and it separates in order: an IgA dimer saturates at four accessible epitopes, an IgA tetramer at eight, an IgM hexamer not until twelve. This is the physical reason IgM excels against repetitive microbial surfaces \cite{Boes2000,Ehrenstein2010} and is nearly useless against small haptens, why secretory IgA is matched to the repetitive glycans of mucosal pathogens, and it follows directly from \cref{eq:matching}: the topology kernel counts only geometrically realisable pairings.

\Cref{fig:ig}(d) places the series on the selectivity axis of \cref{eq:alpha}, using the corona toll and concentration of the worked example of \cref{sec:functional}. The result is worth stating plainly, because it recasts the isotype hierarchy as a physical rather than an immunological fact. Against a threefold difference in epitope density, bivalent IgG achieves $3.0$-fold discrimination, which is exactly the monovalent ceiling of \cref{eq:discrim} and no better: at $\ell=2$ the tuned construct still sits in the saturating regime where discrimination can only equal the density ratio. The IgA dimer reaches $8.8$-fold, the tetramer $47.7$-fold, the IgM pentamer $85.5$-fold and the hexamer $134.7$-fold, the last clearing the monovalent ceiling by a factor of $45$. The per-bond free energy needed to sit at threshold falls in step, from $-16.5\,\kt$ at $\ell=2$ to $-8.4\,\kt$ at $\ell=12$. The isotype series is therefore not a ladder of better antibodies but a traverse of the avidity budget: the same total spend, distributed over more and weaker bonds, converts a detector into a discriminator.

\subsection{The effector face, and why antigen size decides the outcome}

\begin{figure}[t!]
\centering
\includegraphics[width=0.72\textwidth]{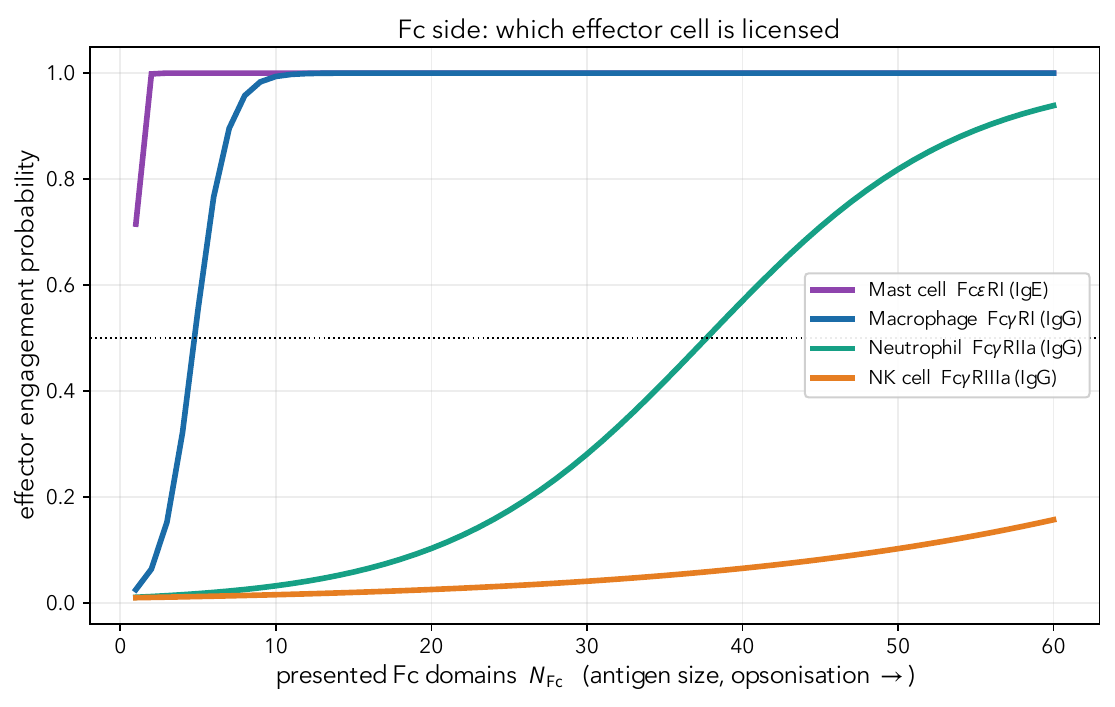}
\caption{The effector face. Engagement probability against the number of presented Fc stems, computed from \cref{eq:compact} with monomeric affinities from Bruhns et al.\ \cite{Bruhns2009}. High-affinity Fc$\varepsilon$RI engages at a single stem; low-affinity Fc$\gamma$RIIIa requires tens, and therefore an immune complex.}
\label{fig:fc}
\end{figure}

Once antibodies coat a target they present an array of Fc stems, and an effector cell reads that array multivalently through its Fc receptors. The number of stems presented, $N_{\rm Fc}$, is set by how large the target is and how heavily it has been opsonised. The affinities of the various Fc receptors differ by four orders of magnitude \cite{Bruhns2009,Bruhns2015}, and when those affinities are put through \cref{eq:compact} a hierarchy of thresholds emerges. \Cref{fig:fc} and \cref{tab:fc} give it.

The result explains a set of otherwise disconnected immunological facts. The mast cell receptor Fc$\varepsilon$RI has an affinity so high that a \emph{single} IgE molecule suffices to arm it, which is why mast cells remain sensitised for weeks by monomeric IgE and why allergy can be triggered by vanishingly small antigen exposures \cite{Kinet1999,Gould2008}. The macrophage receptor Fc$\gamma$RI needs about five stems, an easily met condition on an opsonised bacterium. But the low-affinity receptors, neutrophil Fc$\gamma$RIIa and especially NK cell Fc$\gamma$RIIIa, need tens of stems, and cannot be engaged at all by an antibody bound to a small soluble antigen. This is the physical origin of the requirement for immune complexes in antibody-dependent cellular cytotoxicity \cite{Nimmerjahn2008}. The same logic governs complement: C1q is engaged efficiently only once surface-bound IgG has assembled into ordered hexamers, an explicitly multivalent geometric requirement \cite{Diebolder2014}.

The unifying statement is worth making explicitly, because it is not how immunology is usually taught. \emph{Antigen size does not merely change how well an antibody binds; it selects which effector cell is licensed to respond.} A small antigen permits only high-affinity, low-valency readout. A large one permits the whole repertoire. One geometric parameter, filtered through \cref{eq:compact}, partitions the effector landscape.

\begin{table}[t!]
	\centering
	\small
	\caption{Effector thresholds. $K_A$ is the monomeric affinity \cite{Bruhns2009}, $w_{\rm Fc}=K_Ac_{\rm ref}$ the per-stem weight at an illustrative $c_{\rm ref}=2.5\times10^{-8}$~M, and $N^\star_{\rm Fc}$ the number of stems at which engagement reaches one half.}
	\label{tab:fc}
	\begin{tabular}{@{}llccl@{}}
		\toprule
		\textbf{Effector, receptor} & \textbf{Ig} & $\boldsymbol{K_A}$ [M$^{-1}$] & $\boldsymbol{w_{\rm Fc}}$ & $\boldsymbol{N^\star_{\rm Fc}}$\\
		\midrule
		Mast cell, Fc$\varepsilon$RI   & IgE & $1.0\times10^{10}$ & $250$  & $1$\\
		Macrophage, Fc$\gamma$RI       & IgG & $6.5\times10^{7}$  & $1.6$  & $5$\\
		Neutrophil, Fc$\gamma$RIIa     & IgG & $5.2\times10^{6}$  & $0.13$ & $38$\\
		NK cell, Fc$\gamma$RIIIa       & IgG & $2.0\times10^{6}$  & $0.05$ & $>60$\\
		\bottomrule
	\end{tabular}
\end{table}

\section{Lipoproteins: Valency Amplifies a Single Defect}\label{sec:lipo}

Lipoproteins are multivalent units that most physicists have never thought about and most clinicians think about daily. A lipoprotein is a lipid core wrapped in a shell of amphipathic apolipoproteins, and those apolipoproteins are its binding moieties. Cholesterol traffic in the body is governed by how strongly these particles are captured by hepatic receptors, chiefly the low-density lipoprotein receptor \cite{Brown1986,Goldstein2009}. In the language of \cref{eq:descriptors}, the descriptors that matter are the particle radius $R$, which sets the contact area and hence how many receptors can be reached, and the number and strength of apolipoprotein binding sites.

\begin{figure}[t!]
\centering
\includegraphics[width=\textwidth]{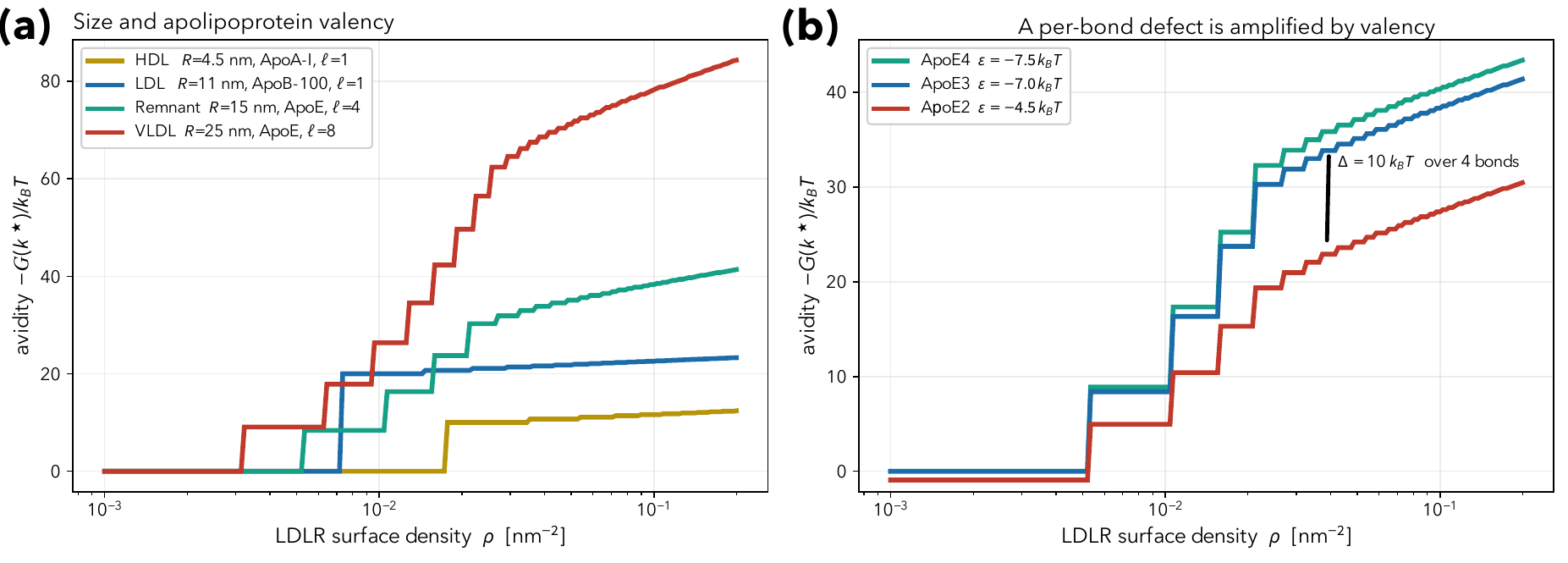}
\caption{Lipoproteins as multivalent units, computed from \cref{eq:withMB} with receptor availability set by the contact area $A_c=2\pi R\delta$ and $\delta=2$~nm. \textbf{(a)}~Avidity against receptor density for four classes. Multivalent ApoE-bearing particles reach far higher avidity than the single-contact strategy of LDL. \textbf{(b)}~Isoform effect for a remnant particle with four ApoE copies. A per-bond change of $2.5\,\kt$ is multiplied by the valency into a $10\,\kt$ shift, a factor of $2\times10^{4}$ in the binding constant.}
\label{fig:lipo}
\end{figure}

Here the framework makes a prediction that is both quantitative and clinically consequential, and it turns on a distinction that is easy to miss. Different lipoprotein classes use fundamentally different strategies for the same receptor. LDL carries a single copy of ApoB-100 \cite{Segrest2001} and makes \emph{one} very strong contact. Remnant and VLDL particles carry several copies of ApoE \cite{Mahley1988,MahleyRall2000} and make \emph{many weaker} contacts. Both work, but they scale differently, and \cref{fig:lipo}(a) shows how. Because avidity grows with valency while a single contact cannot, the multivalent ApoE-rich particles reach far greater avidity: roughly $35\,\kt$ for a remnant particle and $71\,\kt$ for VLDL, against $22\,\kt$ for LDL and only $11\,\kt$ for the small ApoA-I-bearing HDL. Size contributes as well, since a larger particle presents a larger contact area and can reach more receptors, and the two effects compound.

The clinically important consequence appears in \cref{fig:lipo}(b). The human ApoE gene has three common variants, and the ApoE2 isoform carries a substitution that weakens its interaction with the receptor by a large factor \cite{Weisgraber1982,MahleyRall2000}. In our terms this is a change in a single per-bond free energy, $\varepsilon$, of a couple of $\kt$: on its own, a modest effect that would barely register. But the particle carries four copies, and \cref{eq:compact} multiplies the per-bond change by the valency. A $2.5\,\kt$ per-bond defect becomes a $10\,\kt$ deficit in avidity, which is a factor of $2\times10^{4}$ in the binding constant. Clearance collapses, remnant particles accumulate in plasma, and the result is the type III hyperlipoproteinaemia associated with ApoE2 homozygosity \cite{MahleyRall2000,Mahley2009}. Capture in vivo is further staged by an initial low-affinity sequestration on hepatic heparan sulfate proteoglycans before hand-off to the receptor, itself a multivalent step \cite{Williams2008}.

This is the same amplification that makes multivalency powerful, running in reverse. Valency is a lever, and a lever multiplies whatever is applied to it, including damage. It also suggests a general diagnostic principle worth stating: \emph{in a multivalent system, small per-bond perturbations produce large functional consequences, and the amplification factor is the valency.} Anywhere biology uses high valency, it has also created a point of fragility, and mutations at multivalent interfaces should be expected to have outsized effects relative to their structural footprint.

\section{T Cell Recognition and Immune Memory}\label{sec:tcell}

The third example is the one that could not have been treated in Part~I, and it is the reason Part~II was necessary. A T cell must distinguish a peptide derived from a pathogen from the tens of thousands of self peptides displayed on the same major histocompatibility complex molecules, and the discrimination is extraordinary: a single amino acid substitution can convert a fully activating peptide into an inert one \cite{Evavold1991,Davis1998}. The problem is that the binding free energies of agonist and non-agonist peptides differ by only a few $\kt$. At equilibrium, as \cref{eq:discrim} states plainly, a few $\kt$ buys at most one or two decades of discrimination. Observed specificity is far higher. Equilibrium thermodynamics cannot account for T cell recognition, and no amount of refinement of \cref{eq:master} will rescue it.

\begin{figure}[t!]
\centering
\includegraphics[width=\textwidth]{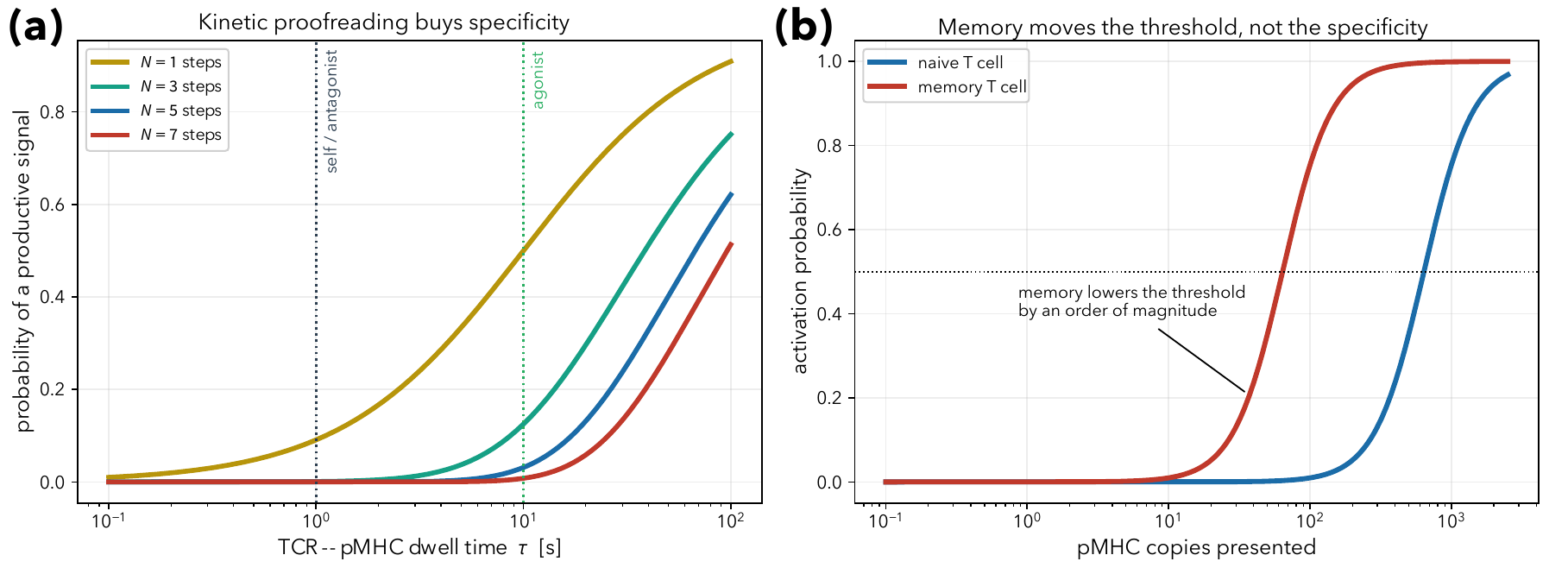}
\caption{T cell recognition as a driven, non-equilibrium process. \textbf{(a)}~Kinetic proofreading, \cref{eq:proof}, with $k_p=0.1$~s$^{-1}$. Increasing the number of proofreading steps sharpens discrimination between a short-lived self complex and a long-lived agonist complex, at the cost of overall signal amplitude. \textbf{(b)}~Naive against memory T cells. Lowering the downstream activation threshold shifts the antigen-density requirement by an order of magnitude while leaving the proofreading machinery, and therefore the specificity, untouched.}
\label{fig:tcell}
\end{figure}

The resolution is the mechanism of \cref{eq:proof}, whose quantitative adequacy for T cells has been established both experimentally and by modelling \cite{AltanBonnet2005,Francois2013}. The T cell does not measure how tightly the complex binds; it measures how long the complex survives \cite{Aleksic2010,Govern2010}, and it does so by requiring the complex to pass through a series of energy-consuming modification steps before it is allowed to signal. \Cref{fig:tcell}(a) shows the effect. With a single step, a tenfold difference in dwell time yields sixfold discrimination, which is roughly the equilibrium result. With three steps it yields $166$-fold, with five $5\times10^3$-fold and with seven $1.5\times10^5$-fold. Specificity is manufactured, not measured, and it is paid for in ATP. That this suffices is remarkable given the numbers: a handful of agonist pMHC among $10^5$ self ligands is enough to trigger a response \cite{Irvine2002,Purbhoo2004}, within a contact whose architecture is itself organised by adhesion receptors into an immunological synapse \cite{Grakoui1999,Dustin2014}.

Notice the trade-off visible in \cref{fig:tcell}(a), because it is not optional. As $N$ increases the curves shift down as well as steepening: the probability that any complex completes the cascade falls. Specificity is bought with sensitivity. A cell that proofreads more stringently needs more antigen to respond at all. This tension is a genuine design constraint, and it sets up the final observation of this Part.

\subsection{What memory changes, and what it does not}

Immune memory is usually described as a faster and stronger response to a previously encountered antigen. The framework lets us ask a sharper question: which parameter changes? There are two possibilities. Memory could improve the \emph{specificity}, by adding proofreading steps or lengthening agonist dwell times, or it could lower the \emph{threshold}, by making the downstream machinery easier to trigger. These are different interventions with different signatures.

\Cref{fig:tcell}(b) shows the second. Lowering the activation threshold, which corresponds biologically to pre-assembled signalling machinery, elevated adhesion receptor density and a relaxed costimulation requirement \cite{Sallusto1999,Farber2014}, shifts the antigen-density response curve by roughly an order of magnitude: a memory cell responds to about $65$ peptide copies where a naive cell needs some $650$. The steepness of the curve, and therefore the discrimination between agonist and self, is unchanged.

This is the resolution of the sensitivity and specificity tension noted above, and it is rather elegant. A naive T cell must be stringent, because it has never seen the antigen and a false positive risks autoimmunity. It therefore sets a high threshold and accepts poor sensitivity. Once an antigen has been validated by a successful primary response, stringency can safely be relaxed for that clone specifically, and the threshold can drop. Memory, in this reading, is not better recognition. It is the same recognition machinery operating with its gate held lower, and the safety of doing so was established during the primary response. The specificity was never the thing that improved.

\section{What the Three Systems Share}

Three systems, one framework. Antibodies show valency setting avidity and geometry gating valency, with the striking consequence that antigen size selects the effector cell. Lipoproteins show the same valency amplification running in reverse, turning a $2.5\,\kt$ per-bond defect into a $2\times10^4$-fold clearance failure. T cells show that when equilibrium discrimination is insufficient, biology breaks detailed balance and buys specificity with metabolic energy, and that immune memory adjusts the threshold rather than the specificity.

Read together they supply a single vocabulary. Valency is a lever, and it amplifies both signal and defect. Geometry decides how much valency is usable. The repulsive layer sets the toll and therefore the threshold. And if the required discrimination exceeds what free energies allow, the only remaining option is to spend energy on a kinetic readout. These are not four separate tricks; they are four faces of \cref{eq:master} and \cref{eq:proof}, and together they determine what any multivalent object, evolved or engineered, is able to discriminate.

\section{Conclusion}\label{sec:conclusion}

We set out to build a theory of biological recognition from the physics of two objects meeting in water, and to carry it through to the point where it explains how living systems actually discriminate.
The foundations turned out to be more constraining than expected. Because everything happens in a liquid, interactions are free energies rather than energies, and solvent averaging manufactures many-body forces whether or not we want them: pairwise additivity fails by almost $20\%$ in a case we could solve exactly, and fails with the opposite sign to the one intuition suggests. Because water screens charge to under a nanometre at physiological salt, electrostatic repulsion cannot hold biological objects apart, so a steric corona is not a refinement but a precondition. And because directionality costs about $4\,\kt$ per constrained partner, single specific contacts are barely worth making, which makes multivalency less a strategy than a requirement.

Adding time did not merely refine these conclusions, it changed their character. Free energies enter rates exponentially, so a tenfold change in receptor number moves the lifetime of a contact from ten milliseconds to months, and thresholds that look gentle thermodynamically are ferocious kinetically. Some discriminations, T cell antigen recognition being the clearest, exceed what any equilibrium mechanism can deliver and are achieved only by breaking detailed balance and paying in metabolic energy. And on the slowest timescale the landscape itself drifts, so that smooth physiological decline produces abrupt functional failure at a bifurcation.

The three biological systems were chosen to be maximally unlike one another, and the fact that one set of equations handles all three is the main evidence that the framework is doing real work. Antibodies showed geometry gating valency, with the consequence that antigen size selects which effector cell may respond. Lipoproteins showed valency amplifying a $2.5\,\kt$ per-bond defect into a $2\times10^4$-fold clearance failure, a reminder that levers multiply damage as readily as signal. T cells showed specificity being manufactured kinetically, and immune memory turning out to adjust the threshold while leaving the specificity untouched.

Running through all three is a single change of question. The quantity that decides a recognition event is not how tightly any one bond forms but how much total avidity is spent, against what repulsive toll, over how many bonds, and in what geometry. Affinity is a property of a ligand; selectivity is a property of an assembly. Once that distinction is taken seriously, several things that look like separate biological phenomena reduce to one piece of physics: the antibody isotype hierarchy, the ApoE clearance defect and the immune memory threshold are all statements about how a fixed avidity is distributed rather than about how large it is.

The framework also has an obvious application that we have deliberately not pursued here. If recognition is a classifier over receptor densities, then a therapeutic construct is an object whose decision boundary can be positioned by design, and the equilibrium and kinetic results of Parts~I and~II become engineering constraints on where that boundary can be placed.
What is missing is mostly data rather than theory. The cooperative couplings collected in \cref{tab:mechanisms} need independent measurement, and any quantitative application needs absolute surface densities of several receptors on the same cells, which is not what current expression data provides. The framework is, in that sense, ahead of its inputs. But it is falsifiable, its parameters are individually measurable, and it makes quantitative predictions that would be straightforward to test on the model systems the field already has.

\appendix
\section{Numerical Conventions}\label{app:numerics}

All partition sums are evaluated in log space by the shift-and-exponentiate identity $\ln\sum_ie^{x_i}=x_{\max}+\ln\sum_ie^{x_i-x_{\max}}$, and bound fractions are obtained as $\theta=\varsigma(\ln z+\ln\Xi)$ with $\varsigma$ the logistic function, which avoids overflow when cooperative couplings make $\ln\Xi$ large. Topology kernels use log-gamma functions rather than factorials. Residence times use the Kramers form \cref{eq:kramers} with an attempt time $\tau_0=1$~ns, appropriate to nanoscale units in water from the diffusivity in \cref{tab:medium}; this prefactor is an estimate and shifts all reported times by a common factor without affecting their ratios.


\end{document}